\documentclass[a4paper,fleqn,usenatbib]{mnras}

\usepackage{newtxtext,newtxmath}
\usepackage{nicefrac,xfrac}

\usepackage[T1]{fontenc}
\usepackage{ae,aecompl}
\usepackage{tikz}

\usepackage{longtable}
\usepackage{pdflscape}

\usepackage{graphicx}	
\usepackage{amsmath}	

\usepackage{orcidlink}

\newcommand{\overbar}[1]{\mkern 1.5mu\overline{\mkern-1.5mu#1\mkern-1.5mu}\mkern 1.5mu}

\title[A search for inter-cluster filaments with LOFAR and eROSITA]{A search for inter-cluster filaments with LOFAR and eROSITA}

\author[D. N. Hoang et al.]{D. N. Hoang$^{1}$\thanks{E-mail: \href{mailto:hoang@hs.uni-hamburg.de}{hoang@hs.uni-hamburg.de}}\orcidlink{0000-0002-8286-646X},
M. Br\"uggen$^{1}$\orcidlink{0000-0002-3369-7735},
X. Zhang$^{2}$\orcidlink{0000-0001-6019-6373},
A. Bonafede$^{3,6}$\orcidlink{0000-0002-5068-4581},
A. Liu$^{2}$\orcidlink{0000-0003-3501-0359},
T. Liu$^{2}$\orcidlink{0000-0002-2941-6734},\newauthor
T. W. Shimwell$^{4,5}$\orcidlink{0000-0001-5648-9069},
A. Botteon$^6$\orcidlink{0000-0002-9325-1567},
G. Brunetti$^6$,
E. Bulbul$^{2}$\orcidlink{0000-0002-7619-5399},
G. Di Gennaro$^1$\orcidlink{0000-0002-8648-8507},\newauthor
S. P. O'Sullivan$^{7}$\orcidlink{0000-0002-3968-3051},
T. Pasini$^1,6$\orcidlink{0000-0002-9711-5554},
H. J. A. R\"ottgering$^5$\orcidlink{0000-0001-8887-2257}, 
T. Vernstrom$^8$,\newauthor
and
R. J. van Weeren$^{5}$\orcidlink{0000-0002-0587-1660}
\\
$^{1}$Hamburger Sternwarte, University of Hamburg, Gojenbergsweg 112, 21029 Hamburg, Germany\\
$^{2}$Max-Planck-Institut f\"ur extraterrestrische Physik,
Giessenbachstra{\ss}e 1, D-85748 Garching, Germany\\
$^{3}$Department of Physics and Astronomy, Bologna University - Via Zamboni, 33, 40126 Bologna, Italy\\
$^{4}$Netherlands Institute for Radio Astronomy (ASTRON), P.O. Box 2, 7990 AA Dwingeloo, The Netherlands\\
$^{5}$Leiden Observatory, Leiden University, PO Box 9513, NL-2300 RA Leiden, the Netherlands\\
$^6$INAF - Istituto di Radioastronomia, 40129 Bologna, Italy \\
$^7$School of Physical Sciences and Centre for Astrophysics \& Relativity, Dublin City University, Glasnevin, D09 W6Y4, Ireland\\
$^8$CSIRO Astronomy \& Space Science, Kensington, Perth 6151, Australia.
}  	
\date{Accepted 2023. Received 2023; in original form 2023}

\pubyear{2023}

\graphicspath{{./}{figs/}}

\begin{document}
\label{firstpage}
\pagerange{\pageref{firstpage}--\pageref{lastpage}}
\maketitle

\begin{abstract}

Cosmological simulations predict the presence of warm hot thermal gas in the cosmic filaments that connect galaxy clusters. This gas is thought to constitute an important part of the missing baryons in the Universe. In addition to the thermal gas, cosmic filaments could contain a population of relativistic particles and magnetic fields. A detection of magnetic fields in filaments can constrain early magnetogenesis in the cosmos. So far, the resulting diffuse synchrotron emission has only been indirectly detected. We present our search for thermal and non-thermal diffuse emission from inter-cluster regions of 106 paired galaxy clusters by stacking the $0.6-2.3$~keV X-ray and 144~MHz radio data obtained with the eROSITA telescope on board the Spectrum-Roentgen-Gamma (SRG) observatory and LOw Frequency ARray (LOFAR), respectively. The stacked data do not show the presence of X-ray and radio diffuse emission in the inter-cluster regions. This could be due to the sensitivity of the data sets and/or the limited number of cluster pairs used in this study. Assuming a constant radio emissivity in the filaments, we find that the mean radio emissivity is not higher than $1.2\times10^{-44}\,{\rm erg \, s^{-1} \, cm^{-3} \, Hz^{-1}}$.
Under equipartition conditions, our upper limit on the mean emissivity translates to an upper limit of $\sim75\,{\rm nG}$ for the mean magnetic field strength in the filaments, depending on the spectral index and the minimum energy cutoff. We discuss the constraint for the magnetic field strength in the context of the models for the formation of magnetic fields in cosmic filaments.
\end{abstract}



\section{Introduction} 
\label{sec:intro}

Cosmological simulations predict the presence of a complex web of cosmic filaments connecting high matter density regions of galaxy clusters. Cosmic filaments are thought to contain, in addition to dark matter and galaxies, warm-hot intergalactic matter (WHIM) with temperatures of $10^5-10^7$~K and low particle densities of $1-10\,{\rm particles\,m^{-3}}$ \citep[e.g.][]{Dave2001}. 

The presence of X-ray diffuse emission in cosmic filaments has been reported by several works over the last few decades. For instance, \cite{Werner2008} reported a $5\sigma$ detection of X-ray emission in the 1.2~Mpc-wide region connecting interacting merging system of Abell~222 and Abell~223 ($z=0.21$) with XMM-Newton $0.5-2.0$~keV observations. 
\cite{Eckert2015} found a filamentary X-ray structure beyond the virial radius ($R_{\rm vir}=2.1$~Mpc) of the massive ($M=1.8\times10^{15}\,M_{\rm sun}$) galaxy cluster Abell~2744 in the XMM-Newton $0.5-1.2$~keV data. 
\cite{Bulbul2016} detected cool (0.8-1~keV) filamentary gas at the distance of $R_{200}$ to the centre of Abell A1750 with \textit{Suzaku}. There is no galaxy cluster found at the location of the filament that is interpreted as the denser, hotter phase of WHIM.
Recently, \cite{Reiprich2021} reported the detection of a 15~Mpc-long region of warm-hot diffuse X-ray emission in between the cluster pair Abell~3391--Abell~3395 and multiple structures beyond $R_{200}$ with the extended ROentgen Survey with an Imaging Telescope Array \citep[eROSITA;][]{Merloni2012,Predehl2021} on board the SRG mission. 
Using indirect stacking technique, \cite{Tanimura2020} found a $4.2\sigma$ X-ray signal from 30--100~Mpc long cosmic filaments \citep{Malavasi2020} using the $0.56-1.21$~keV ROSAT data. 
\cite{Vernstrom2021} reported a $5\sigma$ X-ray diffuse emission in the stacked regions between pairs of luminous red galaxies (LRGs) separated by distance below 15~Mpc using the archival data obtained with the ROSAT All Sky Survey \citep[RASS;][]{Trumper1993}.  
\cite{Tanimura2022} detected excess X-ray emission with $3.8\sigma$ significance from the 463 stacked filaments with length between 30~Mpc and 100~Mpc in the $0.4-2.3$~keV data obtained by eROSITA.

The X-ray emission detected in the bridges from the closely interacting systems (e.g. Abell~222 -- Abell~223, Abell~3391--Abell~3395, part of the LRG pairs in \citealt{Vernstrom2021}) might be due to the heated thermal electrons generated during the merger interaction of the sub-clusters whilst the X-ray structures found in further to the outskirts of galaxy clusters (e.g. beyond $R_{200}$) are likely created by the interaction between the intra-cluster medium (ICM) and its surrounding large-scale structure \citep[i.e. WHIM; e.g.][]{Dolag2006, Werner2008,Planck2013b}  

In the radio band, an indirect detection of radio emission from the regions connecting pairs of LRGs that are thought to trace galaxy clusters/groups has been reported by \cite{Vernstrom2021}. By stacking the GaLactic and Extragalactic All-sky Murchison Widefield Array \citep[GLEAM;][]{Wayth2015,Hurley-Walker2017}  multi-band (73--154~MHz) images, the authors found excess ($5\sigma$) diffuse radio emission in the regions between the LRG pairs. The separation between the LRG pairs is below 15~Mpc with a mean of 9.9~Mpc. However, a following-up analysis by \cite{Hodgson2022} using more sensitive Murchison Widefield Array (MWA) data at 118.5~MHz does not detect the synchrotron signal in between the reported pairs of LRGs. In a recent study, \cite{Vernstrom2023} reports the detection of polarized ($\geq20\%$) radio diffuse emission from the regions between the LRG pairs with the 1.4~GHz Global Magneto-Ionic Medium Survey \citep[GMIMS;][]{Wolleben2021} data and the 30~GHz Planck data \citep{Planck2016}. The LRG pairs are obtained from the Sloan Digital Sky Survey (SDSS) Data Release 7 LRG Catalogue \citep{Lopes2007}. 

Despite of the absence of direct detection of diffuse radio emission from the inter-cluster regions, recent attempts have been made on paired clusters of different separations. A recent attempt by \cite{Locatelli2021} searches for diffuse radio emission between the inter-cluster regions of two pairs of galaxy clusters RXC~J1659--J1702 ($\bar{z}=0.14$) and RXC~J1155--J1156 ($\bar{z}=0.10$) separated by a distance of 14~Mpc and 25~Mpc, respectively, using the LOw Frequency ARray (LOFAR) High Band Antenna (HBA) 8-hours observations. However, no diffuse emission was detected in their $20''$-resolution images that have noise levels of 160 and $260\,{\rm \mu Jy\,beam^{-1}}$. This non-detection suggests that the surface brightness (SB) of the filaments is below the sensitivity of their short observations. 
Similar work without detected signal was done by \cite{Bruggen2021} on the closely (3.1~Mpc) interacting paired clusters Abell~3391 -- Abell~3395 using the Evolutionary Map of the Universe (EMU) Early Science observations \citep[][]{Norris2011} with the Australian Square Kilometre Array Pathfinder \citep[ASKAP;][]{Johnston2007}. 
The strongest evidence for this is the recent direct detection of diffuse radio emission from the connecting regions (namely bridge) between the pairs of clusters (or groups) in Abell~399 -- Abell~401 \citep[3~Mpc][]{Govoni2019,DeJong2022} and Abell~1785N -- Abell~1785S \citep[2~Mpc][]{Botteon2020b}, Coma \citep{Bonafede2022} with LOFAR, and Shapley Supercluster \citep[A3562 -- SC~1329–313;][]{Venturi2022} with the Australian Square Kilometre Array Pathfinder \citep[ASKAP;][]{Hotan2021} and \citep[MeerKAT;][]{Jonas2009}.
However, these regions are still within the virial radii of these clusters \citep{Sakelliou2004a,Botteon2018,Bonafede2022,Venturi2022} and the intra-cluster medium (ICM) of the clusters or groups are in the process of interacting. The resulting shock waves and turbulence can produce diffuse emission in the connecting regions \citep{Govoni2019,Brunetti2020}. Still, these observations are probing further into the cluster outskirts than previously possible and are a stepping stone to directly detect synchrotron emission from cosmic filaments.


Diffuse radio emission in cosmic filaments is expected to be generated by large-scale magnetic fields permeating the source volume and relativistic electrons accelerated by accretion shocks from in-falling matter \citep{Vazza2015a}. Cosmic filaments provide a promising testbed for studying magnetogenesis because the magnetic field within cosmic filaments is expected to be less affected by magnetized outflows from galaxies or dynamo processes than magnetic fields in galaxy clusters. Hence, cosmic filaments provide a good laboratory to study magnetic field amplification by the formation of large-scale structure \citep[e.g.][]{Bruggen2005}. A recent study by \cite{Oei2022} further predicts the radio diffuse signal from the merger- and accretion-shocked synchrotron Cosmic Web (MASSCW) that is generated during the structure formation.

In this study, we search for diffuse radio and X-ray emission from the regions connecting pairs of galaxy clusters that are members of a large-scale structure connecting multiple clusters or groups of galaxies \citep[aka supercluster systems;][]{Bahcall1999}. The extremely faint nature of the inter-cluster filaments makes direct search challenging for the current radio telescopes. Hence, to increase the signal-to-noise ratio (SNR) of the signal we perform stacking of the inter-cluster regions.
Throughout the paper, we assume a $\Lambda$CDM (Lambda cold dark matter) cosmology with $\Omega_m=0.3$, $\Omega_\Lambda=0.7$, and $H_0=70\,{\rm km\,s^{-1}\,Mpc^{-1}}$. 

\section{The supercluster sample} 
\label{sec:sample}

During the performance verification phase of eROSITA, it scans an equatorial sky area of 140 square degrees with its 0.2--10~keV energy band for an exposure duration of 2.2~ks. The observation, namely eROSITA Final Equatorial Depth Survey (eFEDS), reach the sensitivity of the eROSITA All-Sky Survey (eRASS). Detailed analysis of the data by \cite{Liu2022} shows the detection of 542 candidate galaxy clusters in the eFEDS field. \cite{Liu2022} also published the catalog of 19 superclusters with more than 4 members. In this work, we also include the other supercluster systems in the field with less number of cluster members, from two to ten galaxy clusters.

\begin{figure*}
	\centering
	\begin{tikzpicture}
		\draw (0, 0) node[inner sep=0]
		{\includegraphics[width=1\textwidth]{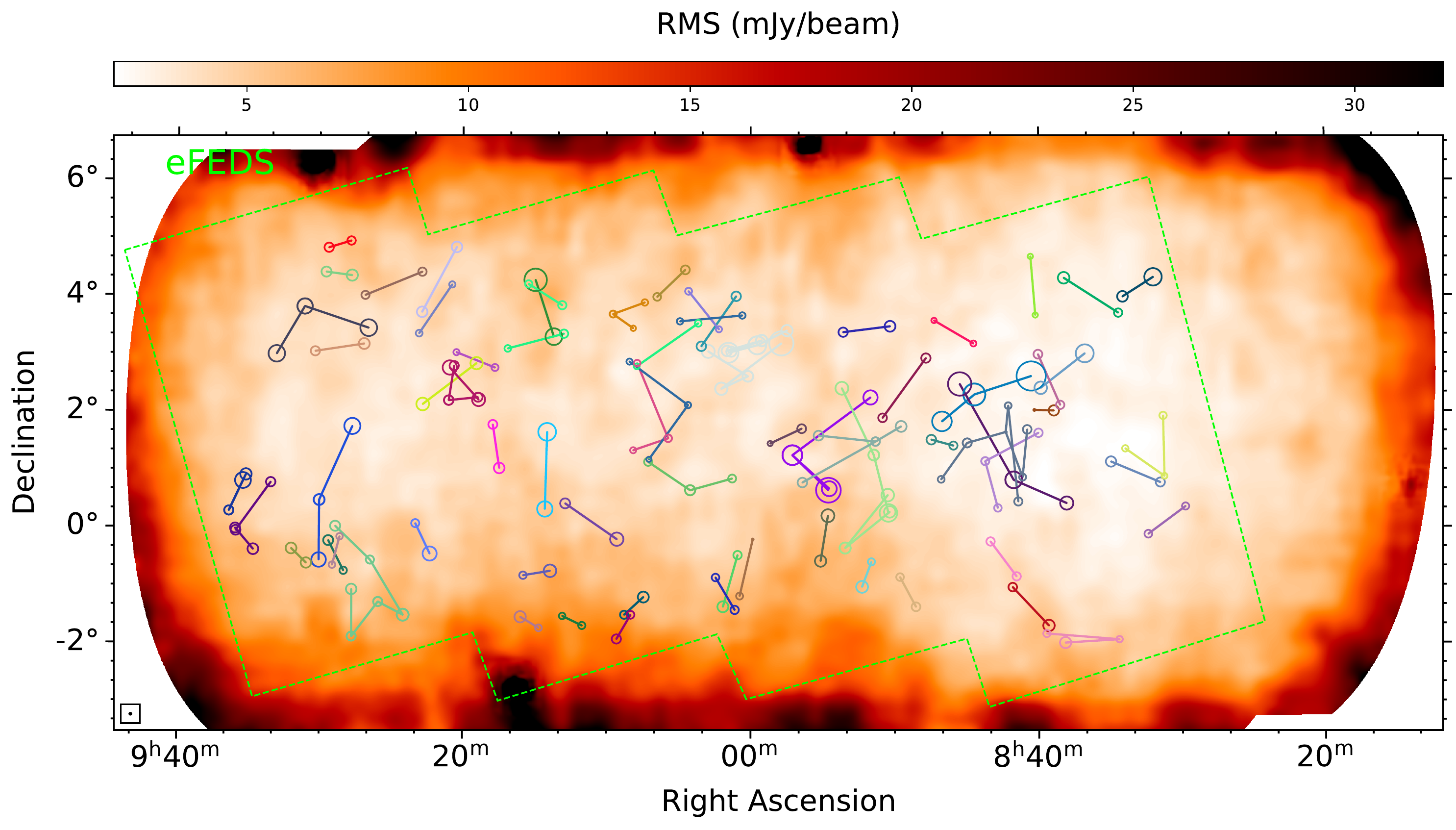}}; 
		\draw (0,-7.) node[inner sep=0]
		{\includegraphics[width=0.33\textwidth]{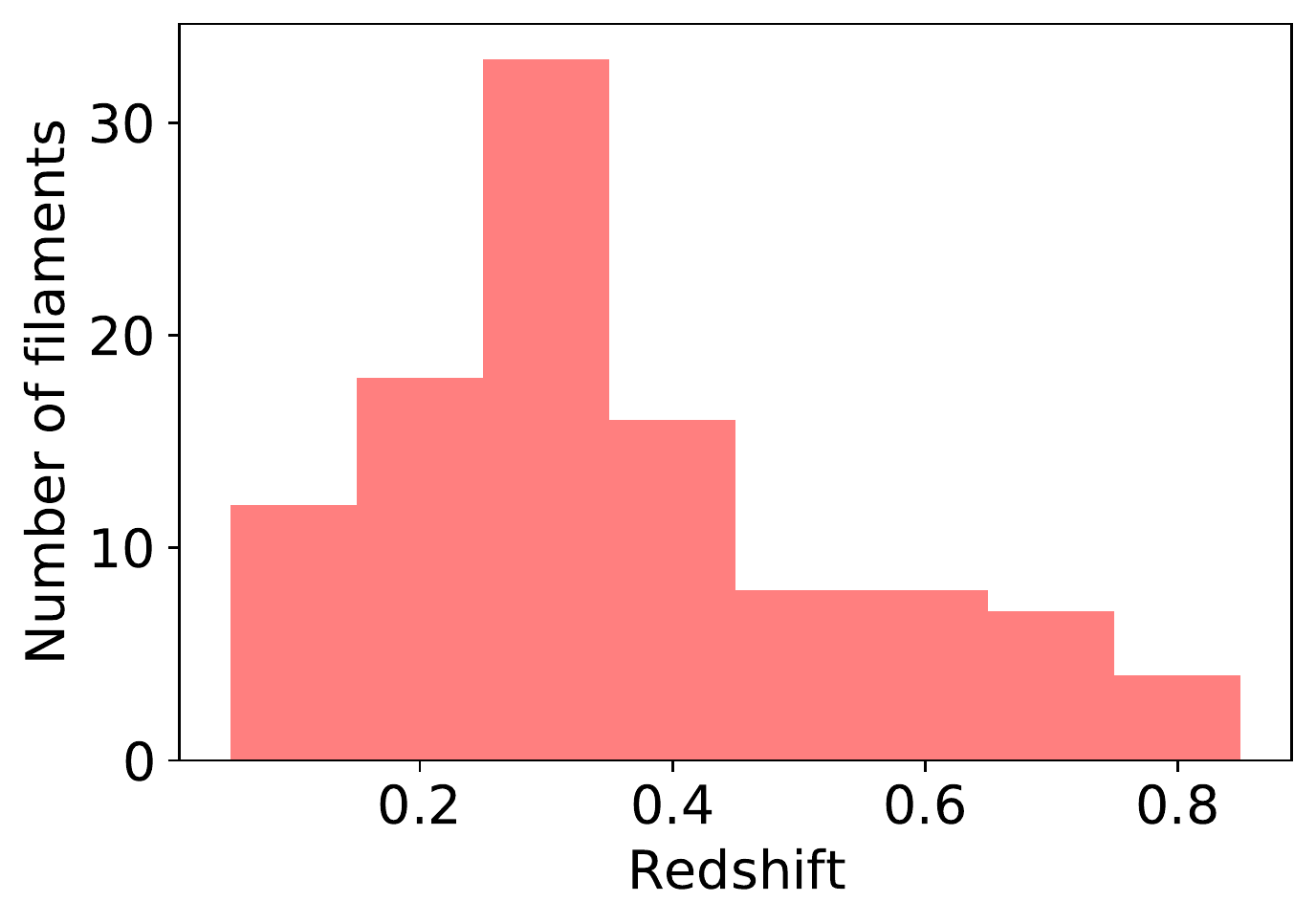}  \hfil
		\includegraphics[width=0.33\textwidth]{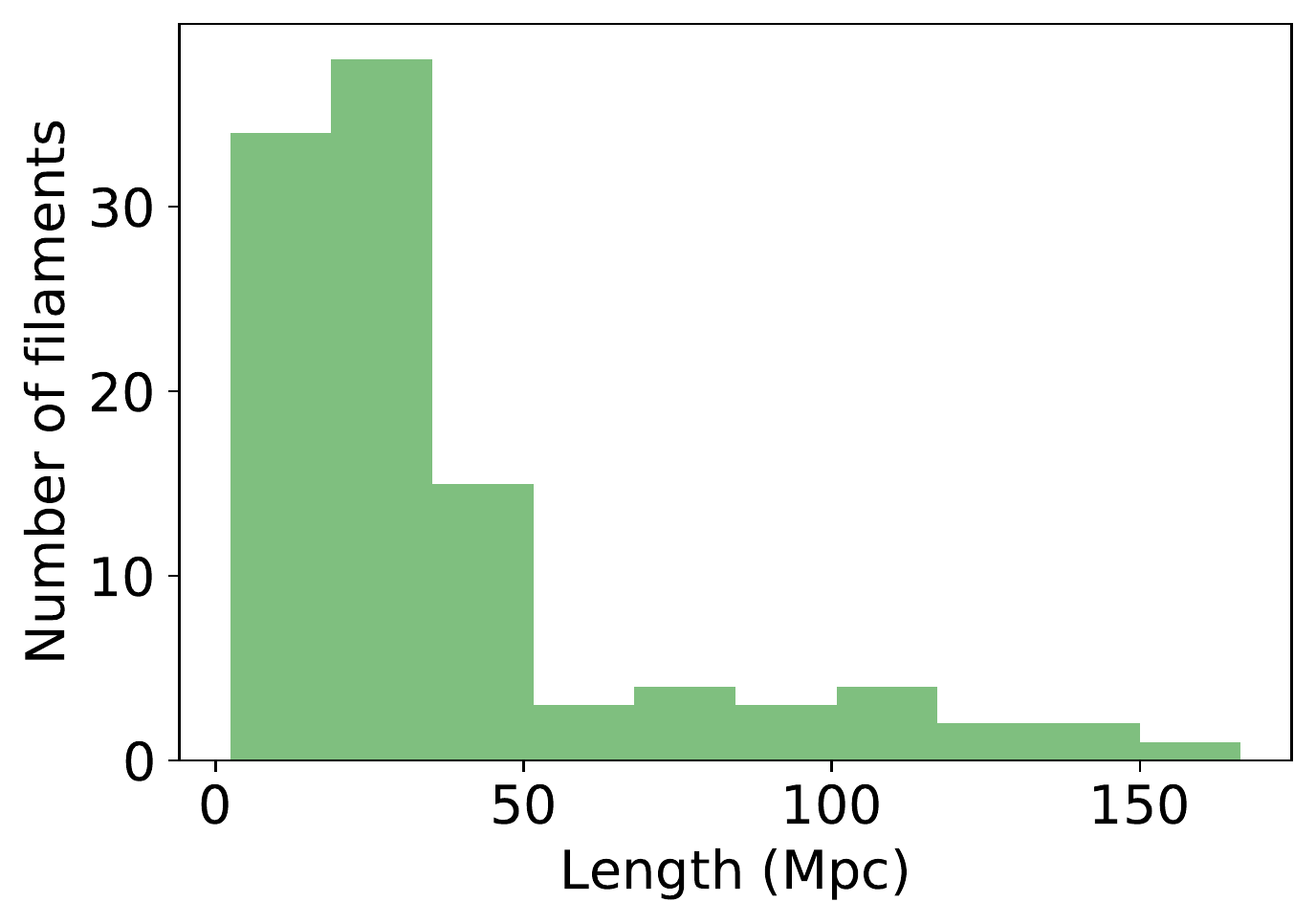}\hfil
		\includegraphics[width=0.33\textwidth]{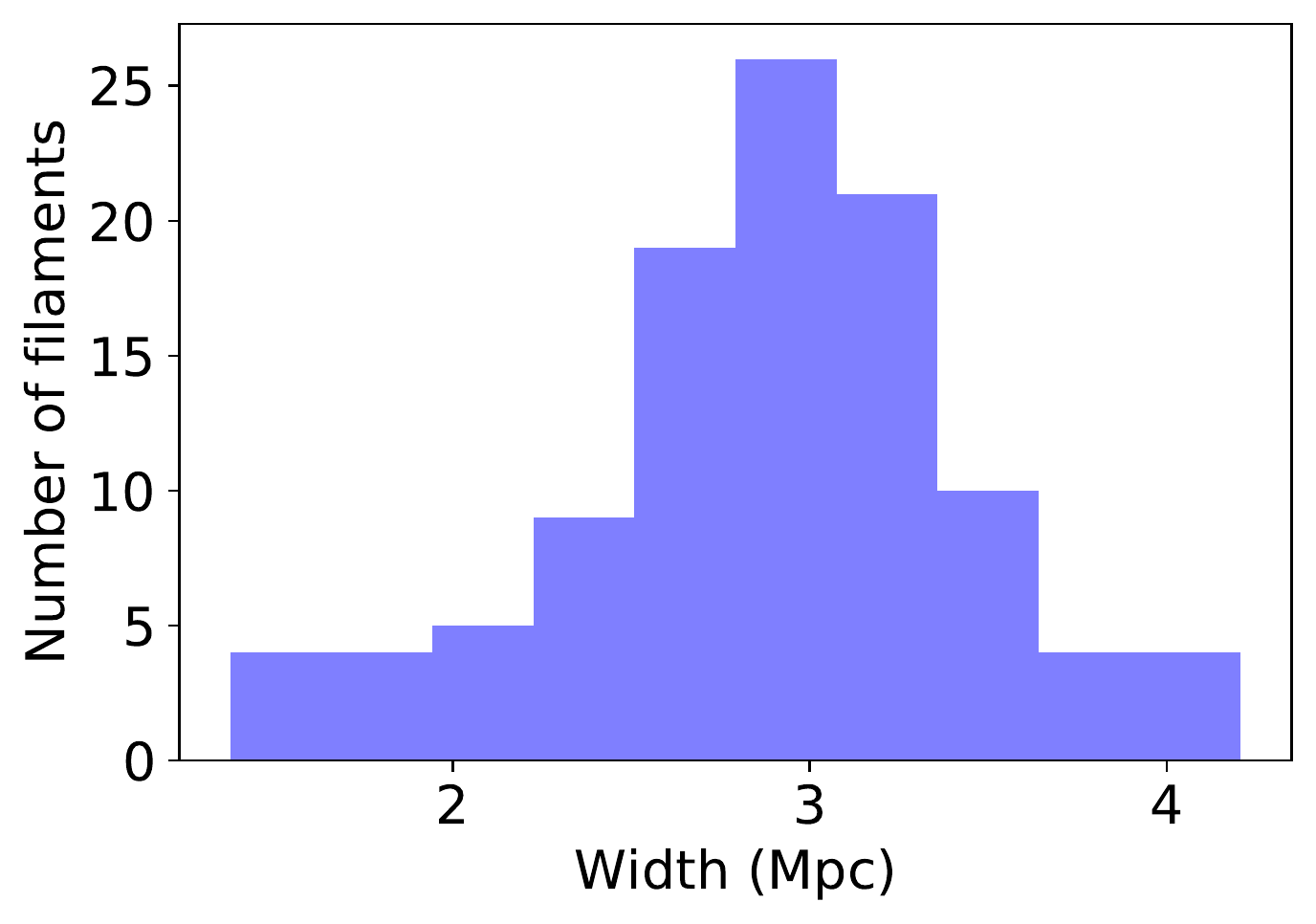} };
		\draw (8.3, 3.1) node {\color{black} (\textit{a})};
		\draw (-3.35, -5.4) node {\color{black} (\textit{b})};
		\draw (2.5, -5.4) node {\color{black} (\textit{c})};
		\draw (8.3, -5.4) node {\color{black} (\textit{d})};
	\end{tikzpicture}
	\caption{(\textit{a}): Spatial distribution of the superclusters in the eFEDS field (green region) on the LOFAR RMS map. Members of a supercluster system are shown with the same colour. The cluster pairs (106) selected in this study are shown with gray lines. The sizes of the circles indicate the relative radius $R_{500}$ of the clusters. (\textit{b}, \textit{c}, \textit{d}): The distributions of the redshift, length, and width of the selected filaments. 
	}
	\label{fig:pairs}
\end{figure*}

We search for diffuse emission from the inter-cluster filaments connecting the supercluster members detected in the eFEDS field. The supercluster catalogue contains the redshifts and sky coordinates of the clusters that allows us to make a list of 162 cluster pairs. These pairs are selected based on the physical distance between the paired clusters in which the closest clusters in a supercluster system are paired together. To maximise the detection of the faint filaments, we select only the cluster pairs that (\textit{i}) are located in the sensitive regions of the LOFAR observations between Dec>$-2^\circ$ and Dec<$+5^\circ$ and (\textit{ii}) are separated with an angular distance from $20\arcmin$ to $120\arcmin$. The final list consists of 106 pairs of galaxy clusters that are shown in Table~\ref{tab:pairs}. Most of these cluster pairs (80 percent) are at redshifts below $0.5$ and are separated from each other by a distance between 4~Mpc and 50~Mpc. The length of a filament is assumed to be the separation of the paired clusters minus their respective virial radii. We define the width of a filament to be two times the mean virial radii of the paired clusters. The virial radius of a cluster is calculated using the scaling relation $R_{\rm vir}=10^{\nicefrac{1}{3}} R_{500}$ \citep{Reiprich2013a}. The radius $R_{500}$ is the radius at which the enclosed mass density is 500 times the critical density of the Universe at the cluster's redshift. For the selected clusters, $R_{500}$ is estimated in \cite{Liu2022}.  In Fig.~\ref{fig:pairs}, we show more detail on the properties of these inter-cluster filaments.

\section{Data}
\label{sec:data}

\subsection{LOFAR radio data}
\label{sec:radio_data}

The eFEDS field was observed with LOFAR for a total duration of 184 hours in multiple projects, including LC13\_029, LT10\_010, LT14\_004, and LT5\_007. The observations make use of 73--75 stations (48 split cores, 14 remotes, and 9--13 international) in full polarisation mode and have a frequency coverage between 120 and 168~MHz. The data was calibrated for direction-independent and direction-dependent effects, following the standard procedures that have been developed for the LOFAR Two-meter-Sky Survey \citep[LoTSS;][]{Shimwell2017,Shimwell2019,Shimwell2022}. The data processing pipelines used for the calibration \texttt{Prefactor}\footnote{\url{https://github.com/lofar-astron/prefactor}} \citep{VanWeeren2016b,Williams2016,deGasperin2019} and \texttt{ddf-pieline}\footnote{\url{https://github.com/mhardcastle/ddf-pipeline}} \citep{Tasse2021} are freely available. Details of the calibration of the eFEDS data were presented in \cite{Pasini2021}. In this study, we combine these calibrated data sets for the eFEDS field, except those obtained from the project LT5\_007 due to its high noise level.

To enhance low surface brightness (SB), large-scale emission from the inter-cluster filaments, we create a low-resolution (2\arcmin) mosaic of the entire field. First, we remove discrete sources to avoid contamination. For each pointing, the subtraction of discrete sources from the data is done using the direction-dependent (DD) calibration solutions and the high-resolution clean-component models that are produced during the calibration steps by \texttt{ddf-pipeline}. After the subtraction, we use the DD-subtracted \textit{uv} data to create wide-field ($8^\circ\times10^\circ$) images for each pointings using \texttt{WSClean}. We set the imaging parameters in such a way that shorter baselines are weighted more for the enhancing of large-scale emission (i.e. \texttt{Briggs'} weighting robust of $0.25$, tapering of long baselines of $10\arcsec$). The \textit{uv} range is set above $45\,\lambda$ which is sensitive to diffuse emission of scales smaller than $1.55^\circ$. The deconvolution is done for the pixels above a threshold of 1.5 times the local RMS, using \texttt{auto-threshold} option. The \texttt{join-channel} deconvolution is applied to take into account the wide bandwidth of the observing frequencies (i.e. 48~MHz). The deconvolved images are corrected for the spatial attenuation of the LOFAR primary beam using the option \texttt{apply-primary-beam} in \texttt{WSClean}.

Before making low-resolution mosaics of the eFEDS field, we smooth the images for each pointing to a common resolution of $2\arcmin$ using \texttt{CASA}'s task (\texttt{imsmooth}). Pixels that are below 10 percent of the LOFAR primary beam sensitivity are blanked out. The smoothed images for each pointing are combined to create a mosaic in a similar procedure as done in \cite{Shimwell2022}. Here the pixels in each images are weighted by the inverse-variance of the local noise  (i.e. $\nicefrac{1}{\sigma^2_{\rm noise}}$) calculated from the pointing images.

\begin{figure*}
	\centering
	\begin{tikzpicture}
		\draw (0, 0) node[inner sep=0]
		{\includegraphics[width=1\textwidth]{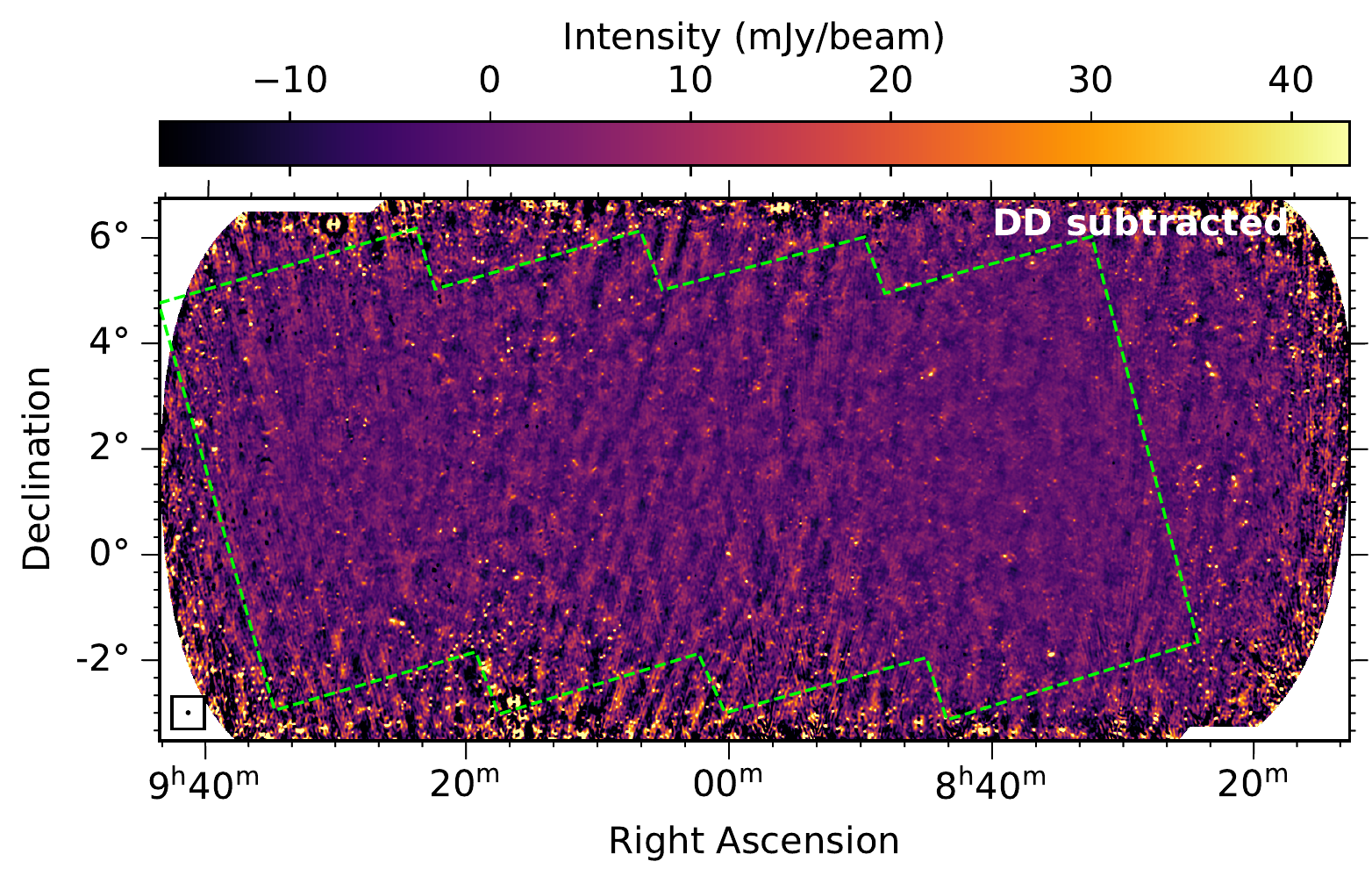}}; 
		\draw (-6.35, 2.7) node {\color{black} \LARGE (a)};
		\draw (0,-10.5) node[inner sep=0]
		{\includegraphics[width=1\textwidth]{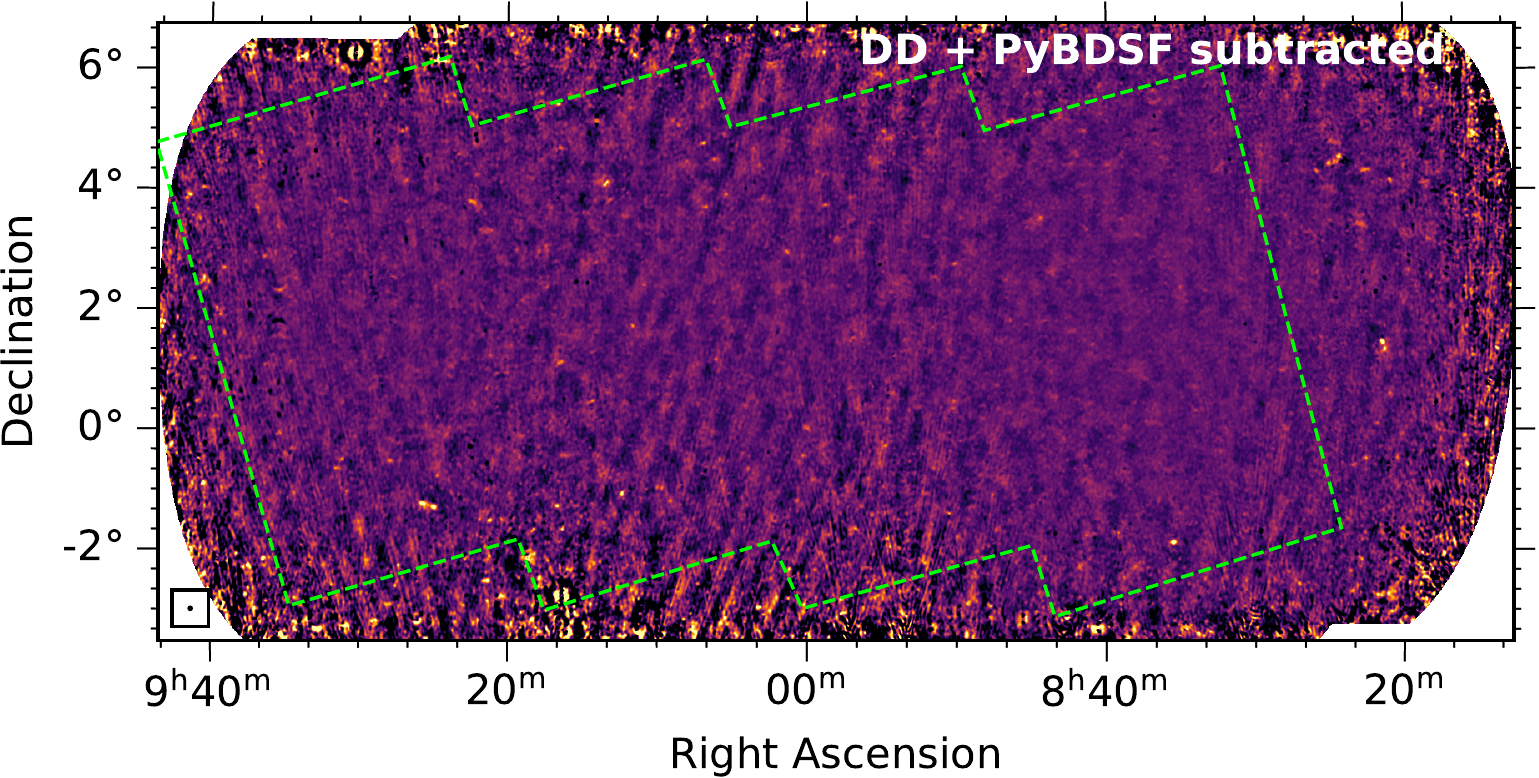} };
		\draw (-6.555, -6.69) node {\color{black} \LARGE (b)};
	\end{tikzpicture}
	\caption{(\textit{a}): Mosaic of the eFEDS field after the subtraction of discrete sources with DD calibration solutions. (\textit{b}): The mosaic after an additional subtraction of discrete sources made with \texttt{PyBDSF}. Both images have the same colour range and field of view. The resolution of the mosaic is shown in the bottom-left corners.
	}
	\label{fig:mosaic_blank}
\end{figure*}

The source-subtracted $2\arcmin$-resolution mosaic of the eFEDS field is shown in Fig.~\ref{fig:mosaic_blank} (a). As seen in the mosaic, majority of the discrete sources are subtracted. However, some residuals are still clearly visible over the field due to the imperfection of the \textit{uv} subtraction. To further remove the residuals of the discrete sources, we use the Python Blob Detector and Source Finder\footnote{\url{https://github.com/lofar-astron/PyBDSF}} \citep[\texttt{PyBDSF;}][]{Mohan2015} to search for the $>3\sigma_{\rm noise}$ pixels that are then fitted with Gaussian functions. To avoid the subtraction from extended emission, only models of the sources that have angular sizes  smaller than three times the beam size are created and are subtracted from the mosaic. The resulting source-subtracted mosaic is shown in Fig.~\ref{fig:mosaic_blank} (b).

\subsection{eROSITA X-ray data}
\label{sec:xray_data}

We use the eFEDS 0.6--2.3~keV count map and exposure map from the early data release\footnote{\url{https://erosita.mpe.mpg.de/edr/eROSITAObservations/Catalogues/liuT/eFEDS_c001_images/}} for data analysis. The detailed data reduction procedure was described in \cite{Brunner2022}. The image products are rebinned to a pixel size of 16\arcsec$\times$16\arcsec. Extended sources in galaxy clusters and groups catalog \citep{Liu2022} are masked using a radius of $3\times$\texttt{R\_SNR\_MAX}. Point sources in the catalog of \cite{Brunner2022} are masked using $1.5\arcmin$-radius masks. We visually inspect the source masked image and apply additional masks to bright source PSF wings that exceed the $1.5\arcmin$-radius. We do not use the 0.2--0.6 keV band image because the foreground emission from Galactic halo is dominant in this band \citep{Ponti2022}, which is also the possible reason why \citet{Tanimura2022} only reported  filament emission detection using the 0.6--2.3 keV images. 

\section{The stacking technique} 
\label{sec:stacking}

Diffuse emission from cosmic filaments is expected to be faint due to its low particle density. To search for the diffuse emission from these inter-cluster regions, previous studies exploited the stacking technique of multiple individual images to increase the SNR of the sources  \citep[e.g.][]{Clampitt2016,Tanimura2020,Tanimura2022,Vernstrom2021,Vernstrom2023}. 

\subsection{Stacking of radio data}
\label{sec:stacking_radio}

We combine the radio images round the cluster pairs that is roughly similar to the procedure described in \cite{Clampitt2016} and \cite{Vernstrom2021}.
For each pair of galaxy clusters, cutout images are extracted from the $2\arcmin$-resolution mosaic (see Fig~\ref{fig:mosaic_blank}, b). As the cluster pairs have different orientations in the sky, we rotate the cutout images to align the cluster pairs along the horizontal ($X$) axis of the image. The angular sizes of the cutout images in the $X$ and $Y$ axes are set to common angular sizes in the units of the cluster separations ($d$) and the mean of the virial radii ($\overbar{R}_{\rm vir}=\nicefrac{(R_{\rm vir,1}+R_{\rm vir,2})}{2}$) of the paired clusters, respectively. The cutout images are then regridded to common pixel sizes (i.e. $1024\times1024$ pixel$^2$). The centres of each paired clusters are located on the same pixel locations in the regridded images (i.e. $(-256, 0)$ and $(+256, 0)$, here the coordinate origin is at the cutout image centre). The pixel sizes in $X$ and $Y$ axes are set to be \nicefrac{$d$}{512} and \nicefrac{1024}{(10$\overbar{R}_{\rm vir}$)}, respectively. When the cutout images are regridded, the images are stretched or contracted, depending on the original angular separation of the paired clusters. We convert the SB unit to Jy~pixel$^{-1}$ during this step as the SB unit of Jy~beam$^{-1}$ is not relevant. During the rotation and regridding, the flux density is conserved.

The cutout images with common pixel sizes are combined to generate a stacked image of the cluster pairs. Individual pixels are averaged and are weighted by the inverse-variance ($\nicefrac{1}{\sigma^{2}_{\rm noise}}$) of the noise estimated from the cutout images.

To estimate the volume emissivity of the plasma filling a filament we assume a tube-like shape for the filament. The area of the filament cross-section is $A=\pi r^2$, where $r$ is the radius of the cross section of the filament. The emissivity and integrated flux density\footnote{The synchrotron spectrum convention of $S\propto\nu^{-\alpha}$ is used in this study.} for the filament are
\begin{equation} 
\label{eq:eps}
	\begin{split}
		\epsilon & = \frac{P}{V}, \\
		S &= \frac{P}{4\pi D_L^2 K(z)} = \frac{\epsilon l r^2}{4K(z) D_L^2},
	\end{split}
\end{equation}
where $P$ is the radio power, $V$ is the filament volume, $l$ the length of the filament, and $D_L=\nicefrac{(D_{L,a}+D_{L,b})}{2}$ is the luminosity distance to the pair clusters (the subscripts a and b denote the two clusters of the pair), $K(z)=(1+z)^{(\alpha-1)}$ is the monochromatic \textit{K}-correction term for a radio source of spectral index $\alpha$ at redshift $z$. When a number of $N$ filaments are stacked, the mean flux density for the filaments is,
\begin{equation} 
\label{eq:S_mean}
	S = \sum_{i=1}^{N} \omega_i S_i = \sum_{i=1}^{N} \frac{\omega_i \epsilon_i l_i r_i^2 }{4 K(z_i) D_{L,\,i}^2},
\end{equation}
where $\omega_i=\frac{\nicefrac{1}{\sigma_{\rm noise, \textit{i}}^2}}{\sum_{j=1}^{N}{\nicefrac{1}{\sigma_{\rm noise, \textit{j}}^2}}}$ is the inverse-variance weighting of the cutout images. Assuming that the emissivity of the filaments is constant for all filaments in the sample, we can write equation~\ref{eq:S_mean} as follows:
\begin{equation} 
\label{eq:eps_mean}
	\epsilon_0 = \frac{S} {c},
\end{equation}
where $c= \sum_{i=1}^{N} \frac{\omega_i l_i r_i^2 }{4K(z_i) D_{L,\,i}^2}$. The mean flux density, $S$, is measured from the stacked image and $c$ is calculated for $N$ cluster pairs.

\subsection{Stacking of X-ray data}
\label{sec:stacking_xray}

For the X-ray data, our stacking method is similar to that of \citet{Tanimura2022}. For each cluster pair, we extract the SB 1D profile from the connecting segment to $5\times\bar{R}_\mathrm{vir}$ with bin sizes of $0.5\bar{R}_\mathrm{vir}$ (e.g. see Fig.~2 in \citealt{Tanimura2022}). The SB of the $i$th bin is 
\begin{equation}
    SB_i = \frac{\sum_j N_{ij}}{ \sum_{j} E_{ij}\times A_{ij} },
\end{equation}
where $N_{ij}$, $E_{ij}$ and $A_{ij}$ are the count number, vignetting corrected exposure time and sky area of the $j$-th  pixel in the $i$-th bin. 
For the 106 extracted profiles, we use a bootstrapping method to evaluate the mean and $1\sigma$ uncertainty of the combined profile. Note that the ratio between the vignetting corrected and uncorrected exposure maps is flat across eFEDS, we do not distinguish instrumental background and sky X-ray background. Instead, we will use the SB outside $\bar{R}_\mathrm{vir}$ as the local background.

\section{Injection of filament models}
\label{sec:model_inj}

\begin{figure}
	\centering
	\includegraphics[width=1\columnwidth]{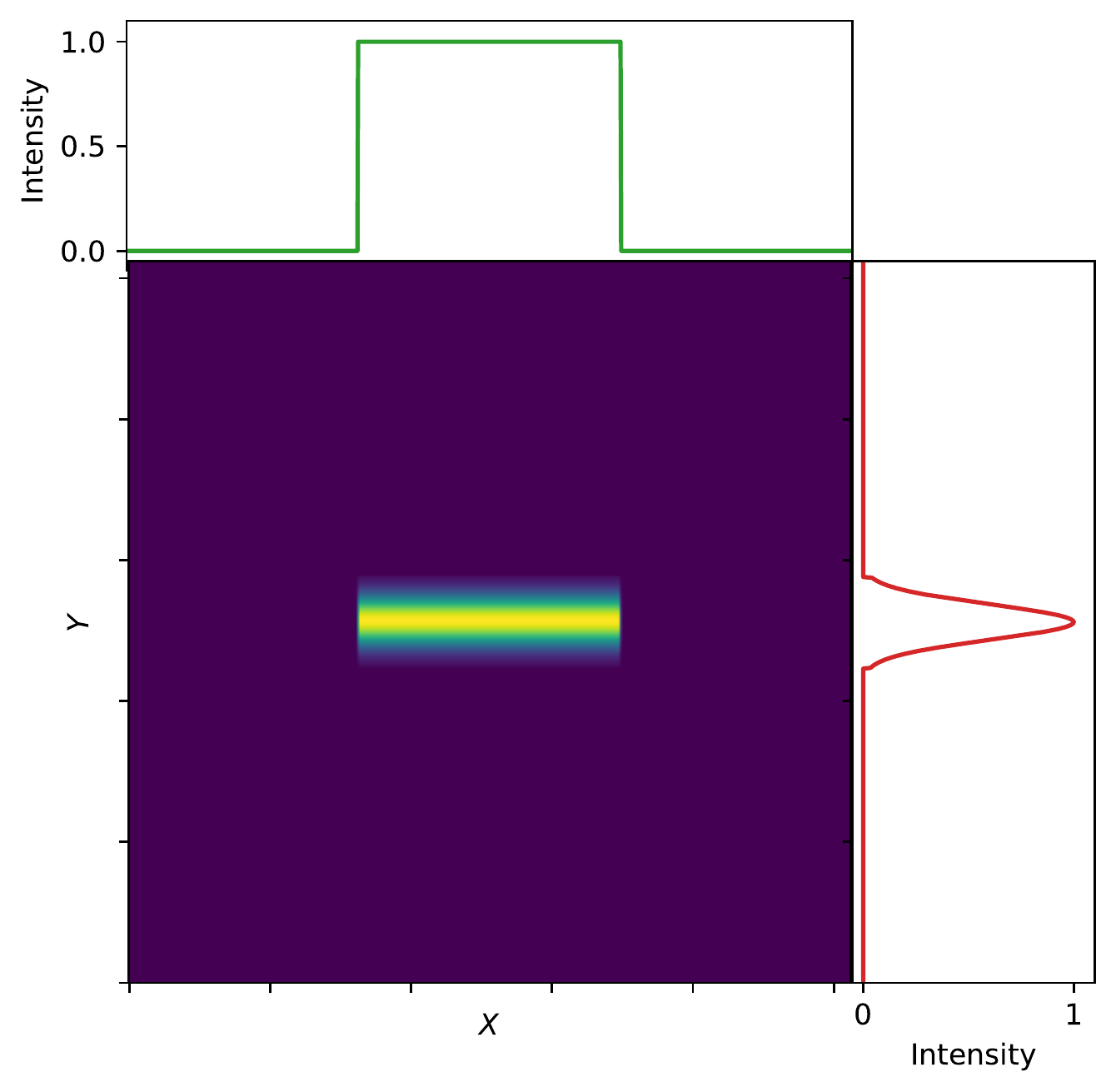} \\
	\caption{Filament model and its profiles across the width (right sub-panel) and length (top sub-panel).
	}
	\label{fig:model}
\end{figure}

Due to the missing of \textit{uv} sampling, radio interferometric observations suffer from recovering large-scale, low-SB emission. A recent study by \cite{Bruno2023} extensively addressed this issue with LOFAR data for cluster-scale diffuse emission. They found that the  LOFAR observations miss only 10 percent of the flux density from diffuse sources of $10.5\arcmin$ scales. However, the missing flux density can be up to 50 percent for $18\arcmin$-scale sources.
Our LOFAR observations might miss part of diffuse emission from the inter-cluster filaments that span a large-spatial area of the sky, up to $\sim$1.5 degree in size. We examine this effect in the LOFAR data by the injection of filament models and compare the flux density before and after the injection. We assume a simple tube-like 3D shape for the filaments. In projection, the width $w$ and length $l$ of a filament are defined as
\begin{equation} 
\label{eq:wl}
	\begin{split}
		w & = 2 \overbar{R}_{\rm vir}, \\
		l &= d-R_{\rm vir,1}-R_{\rm vir,2}.
	\end{split}
\end{equation}
The projected SB across the filament width is set to follow a Gaussian function with a standard deviation of $\sigma_{\rm Gaus.}=\nicefrac{\overbar{R}_{\rm vir}}{3}$. With this SB distribution, the amount of the flux density integrated over the filament width (i.e. between $-3\sigma_{\rm Gaus.}$ and $3\sigma_{\rm Gaus.}$) is 99.7 percent. Along the length of the filaments, the projected SB is assumed to be constant. An example of a filament model is shown in Fig.~\ref{fig:model}.

For each pointings, the models of the filaments are multiplied with the corresponding LOFAR primary beam to account for the spatial sensitivity of the telescope. The primary-beam-attenuated models for the filaments are then performed Fourier transformation and are added (injected) to the DD-subtracted \textit{uv} data in which discrete sources have been removed by the DD-calibration solutions and the \texttt{ddf-pipeline} high-resolution ($\sim6\arcsec$) models. The injected data sets are then processed in the similar manner as done in Sec.~\ref{sec:radio_data} to generate a final mosaic of the eFEDS field. Due to the low SB of the mock filaments, the injected emission is not deconvolved during the imaging. To avoid the removal of the injected filaments in the \texttt{PyBDSF}-subtracted step, we use the models for the residuals which are created with the \texttt{PyBDSF} run on the original mosaic (i.e. Fig.~\ref{fig:mosaic_blank}, \textit{a}). As mentioned in Sect.~\ref{sec:radio_data}, we only made the models for the residuals with an angular size smaller than three times the synthesized beam to avoid the removal of the large-scale signal from inter-cluster filaments. We refer to this injection of the filament models in to the \textit{uv} data as the \textit{uv} injection.

The flux density of the filaments might be lost during the stacking due to the image processing. We also test this possibility by injecting of the filament models directly in to the cutout images, which is referred as the \textit{image} injection. The models of the filaments are smoothed to the same $2\arcmin$-resolution of the cutout images before being added to the cutout images. The model-injected cutout images are then combined to generate stacked images, following the procedure described in Sec.~\ref{sec:stacking}.

\section{Results} 
\label{sec:res}

\subsection{Radio stacked image}
\label{sec:res_radio}

\begin{figure*}
	\centering
	\includegraphics[width=0.8\textwidth]{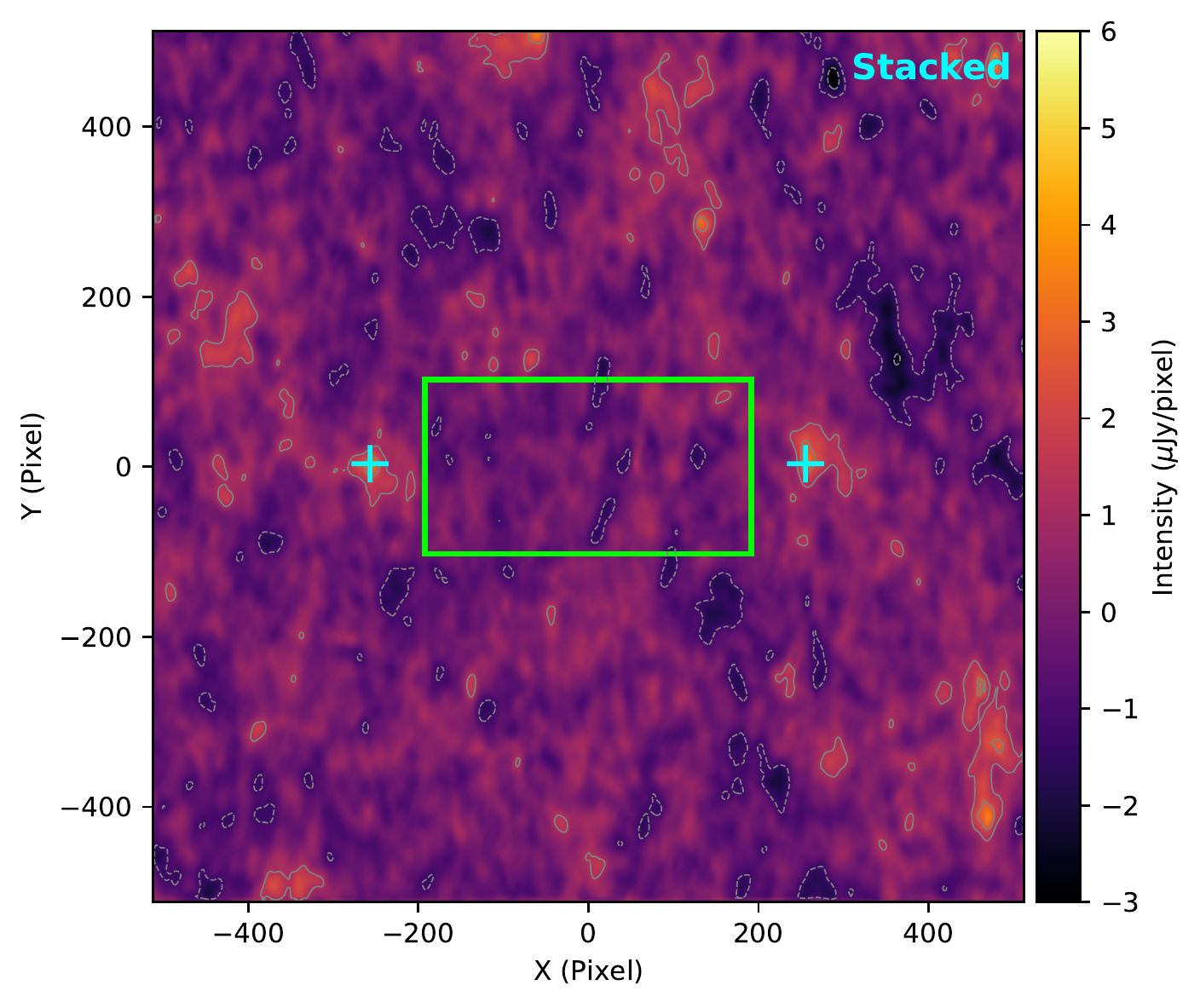} \\
	\caption{LOFAR~144~MHz stacked image of the 106 cluster pairs. The locations of the paired clusters are marked with plus (+) signs. The width and length of the green rectangle region is $w$ and the mean of the filament lengths, respectively. The contours are drawn at $[\pm2, \pm4,\pm6]\times\sigma_{\rm noise}$, where $\sigma_{\rm noise}=0.6\,{\rm \mu Jy\,pixel^{-1}}$. Dashed lines represent negative contours. 
	}
	\label{fig:stack}
\end{figure*}

In Fig.~\ref{fig:stack}, we show the stacked image for the 106 selected pairs of galaxy clusters. The sensitivity of the stacked image is $\sigma_{\rm noise}=0.6\,{\rm \mu Jy\,pixel^{-1}}$, where the pixel sizes along X and Y axes are in the units of $\nicefrac{d}{512}$ and $\nicefrac{1024}{(10\overbar{R}_{vir})}$. At the locations of the paired clusters, radio emission above $2\sigma_{\rm noise}$ is seen. However, no diffuse emission in the regions connecting the paired clusters is detected. In the following subsections, we use the stacked image to constrain on the upper limits of the flux density, the emissivity, and the magnetic field strength for the filaments.

\subsubsection{Flux density}
\label{sec:flux}

We constrain the upper limit of the mean flux density for the inter-cluster regions to be $S < 3 \Delta S$, where $\Delta S=\sqrt{N} \Delta S_{\rm sub}$ and $\Delta S_{\rm sub}=n \sigma_{\rm noise}$ are the uncertainties of the flux densities over the filament regions and independent sub-regions, respectively; $N$ is the mean number of the independent sub-regions consisting of $n$ pixels each. The factor of 3 here indicates that the true flux density of the filaments is below the upper limit $S$ with a probability of 99.7 percent.  
As defined in Eq.~\ref{eq:wl}, the filament widths are equal to two times the mean of the virial radii of the paired clusters, and that the filament lengths span between the boundaries of the virial radii of the clusters. The area of the filaments ranges from 10 to 236 independent beams with a mean of 51 beams. Almost 85 percent of the filaments have an area smaller than 80 beams. More detail on the distribution of the filament area is shown in Fig.~\ref{fig:fil_hist} (\textit{a}). 
When the number of independent sub-regions is taken to be the mean of the filament area in beam unit ($N=51$), the uncertainty of the flux density in an independent sub-region of the filaments is $\Delta S_{\rm sub}=0.9\,{\rm mJy}$, here $n=1540\,{\rm pixels}$. This results in the mean flux density of the inter-cluster regions to be lower than $S<19.3 \,{\rm mJy}$. The upper limit for the flux density is estimated here under the assumption that the SB is constant over the whole filament area. Hence, this upper limit is likely much higher than the true flux density for the filaments as the SB in the outer regions is expected to be fainter than the SB in the middle regions (along the length of the filaments).

\subsubsection{Emissivity}
\label{sec:emissivity}

\begin{figure*}
\centering
\begin{tikzpicture}
\draw (0, 0) node[inner sep=0] {
    \includegraphics[width=0.19\textwidth]{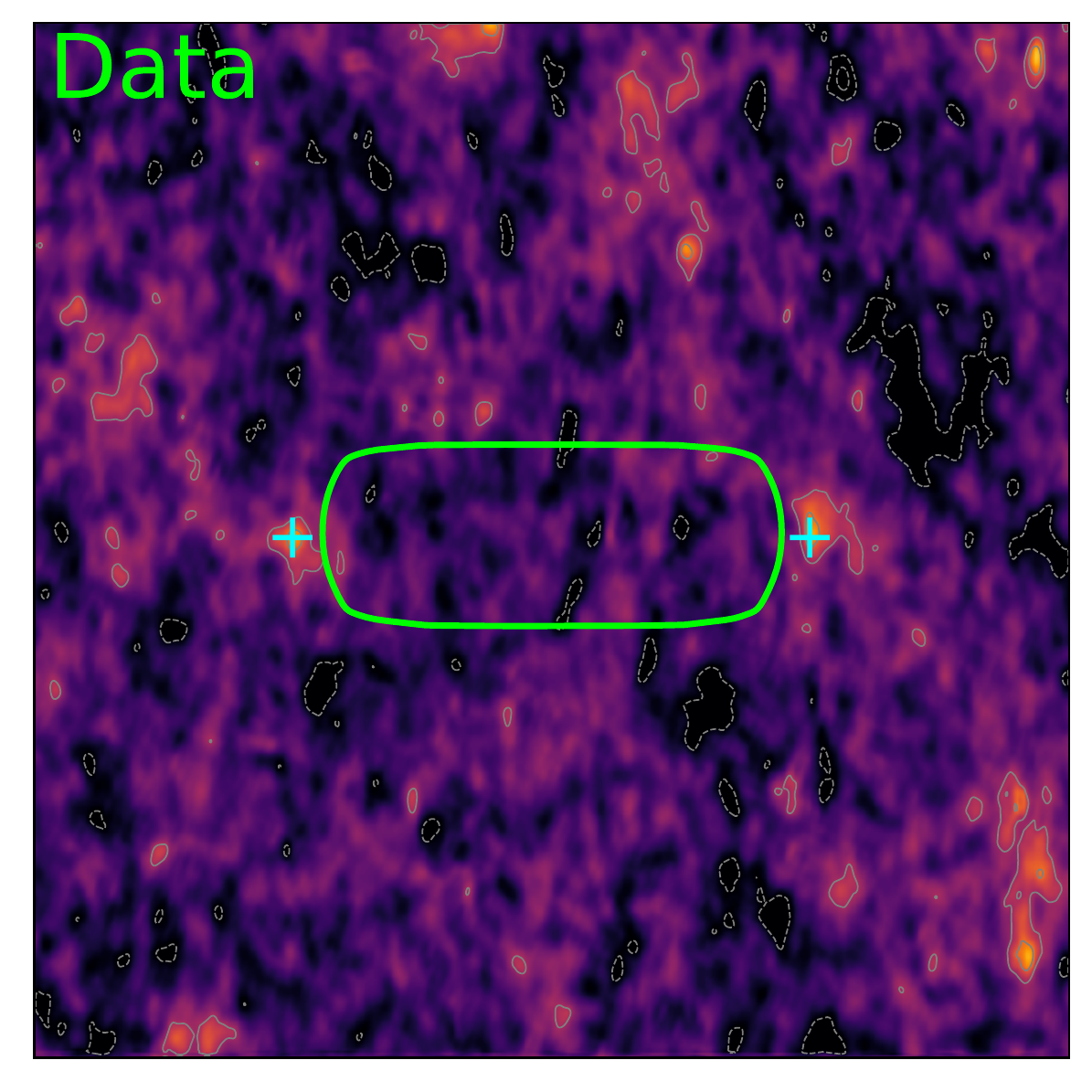}  \hfil
    \includegraphics[width=0.19\textwidth]{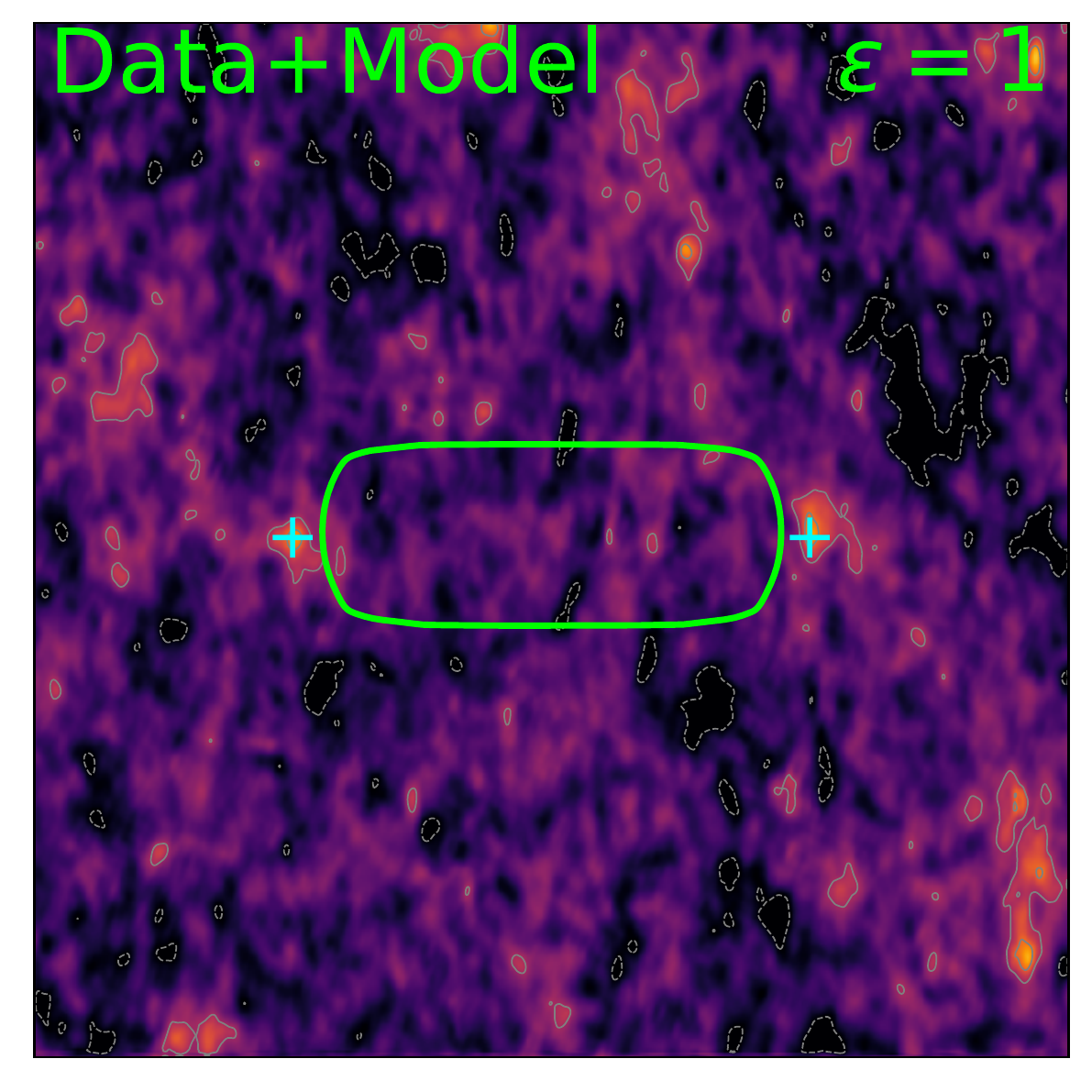} \hfil
    \includegraphics[width=0.19\textwidth]{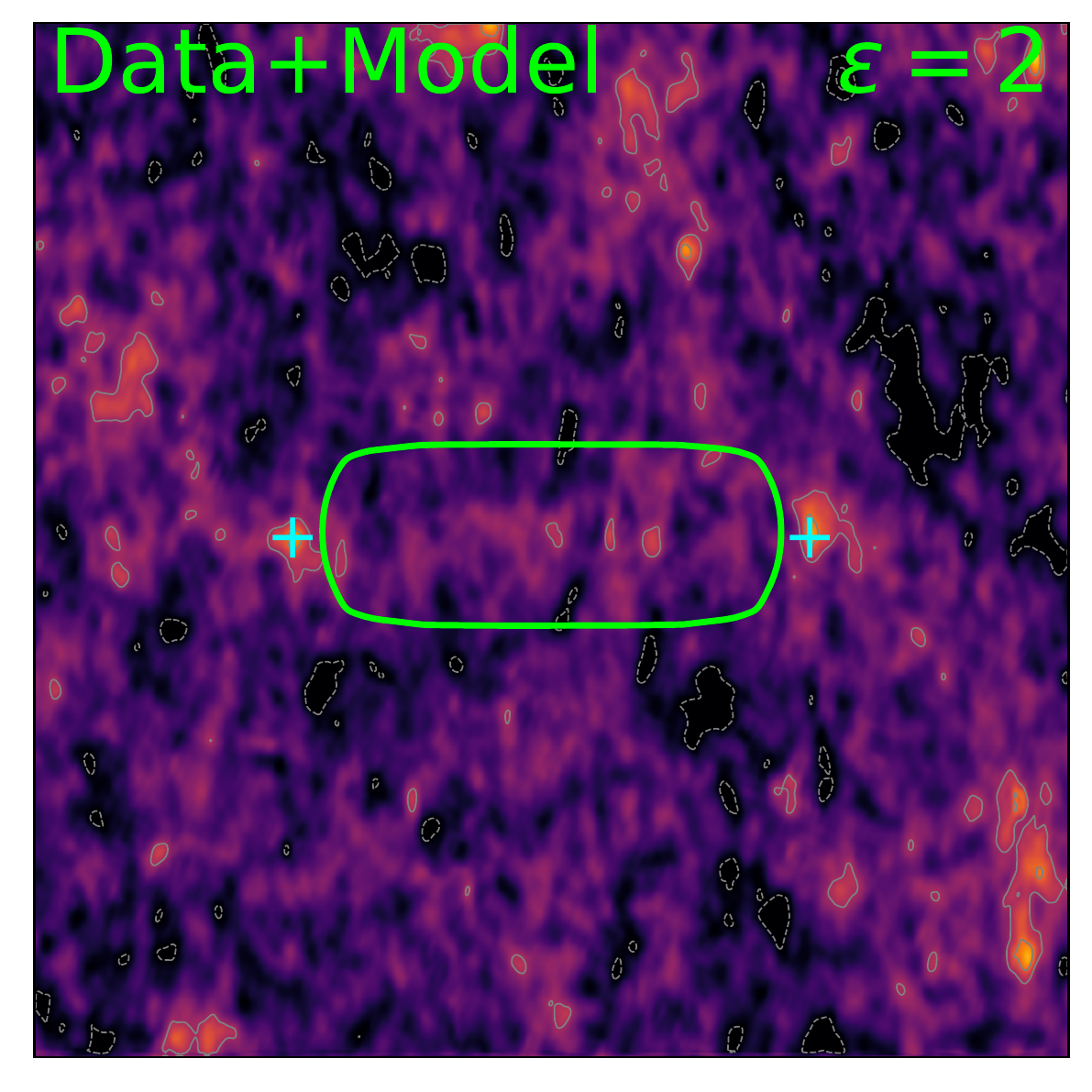} \hfil
    \includegraphics[width=0.19\textwidth]{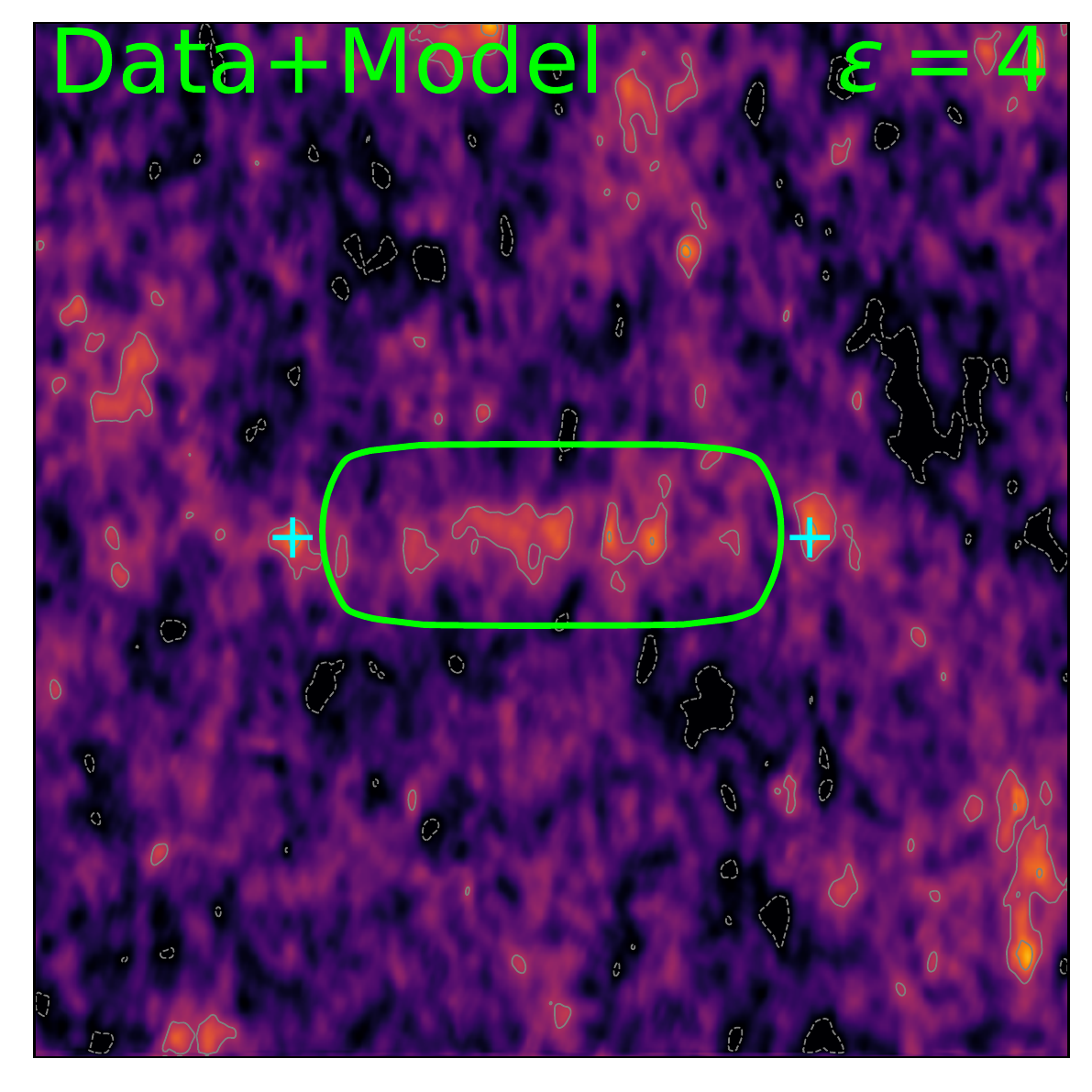} \hfil  
    \includegraphics[width=0.19\textwidth]{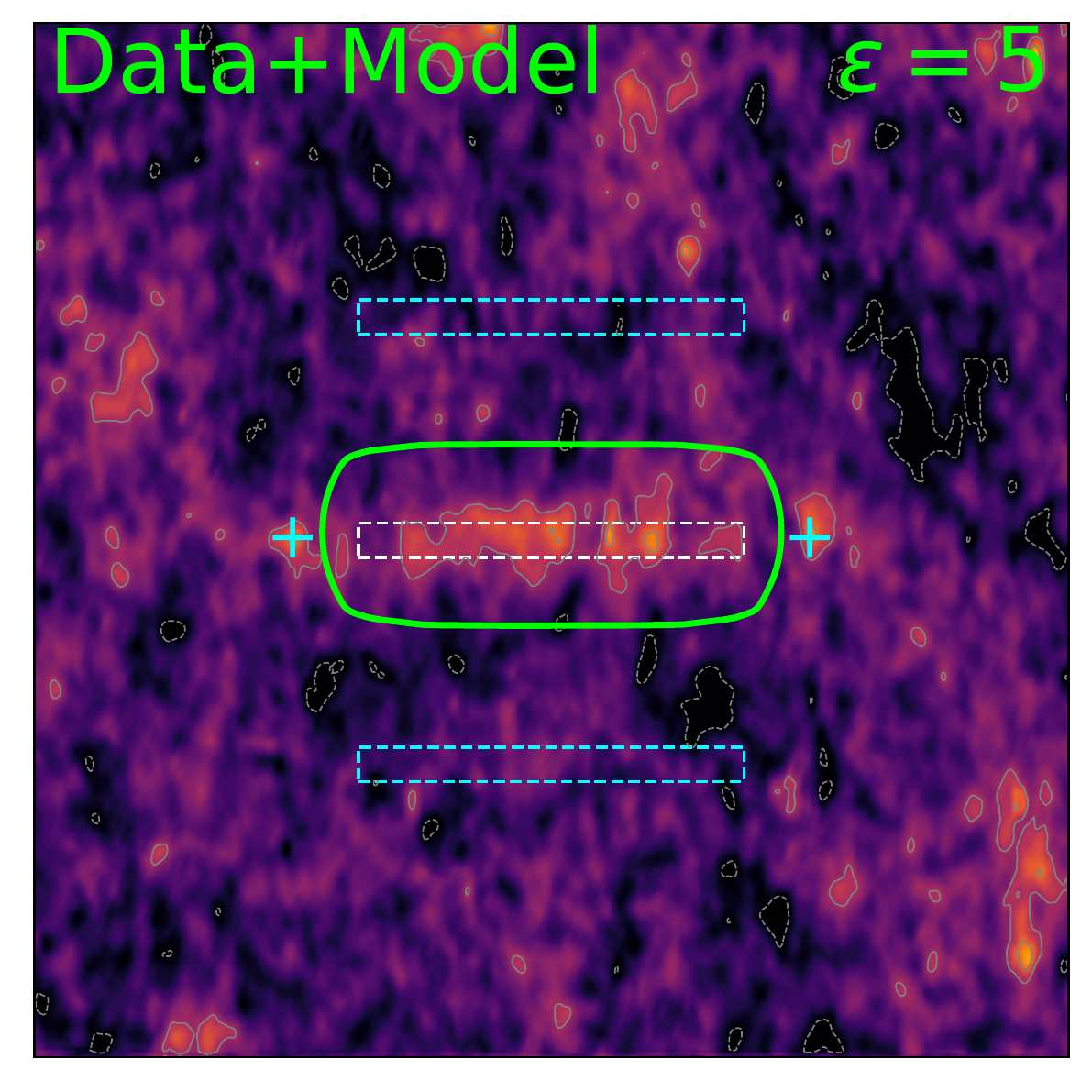}    };
\draw (0,-3.5) node[inner sep=0] {
    \includegraphics[width=0.19\textwidth]{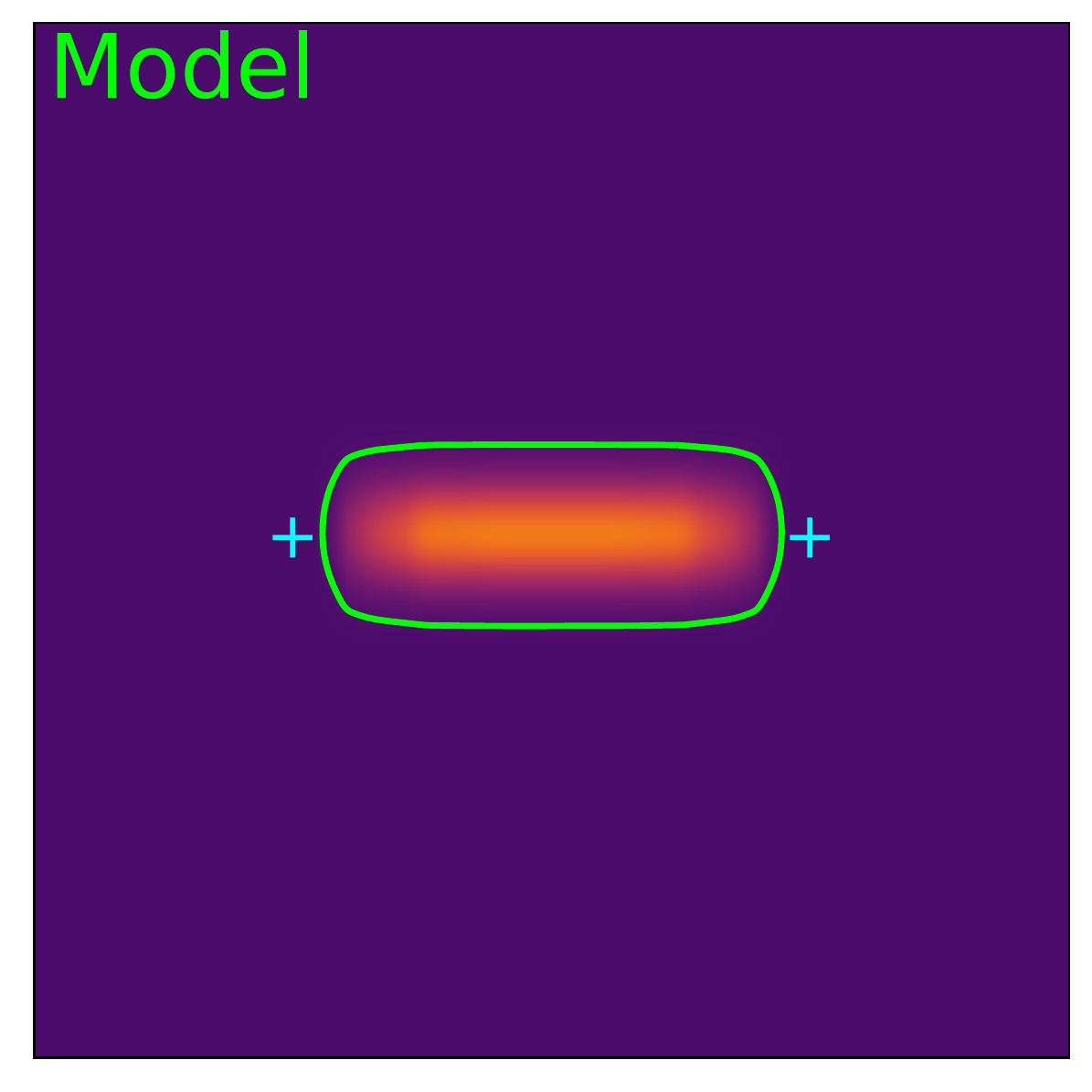}  \hfil
    \includegraphics[width=0.19\textwidth]{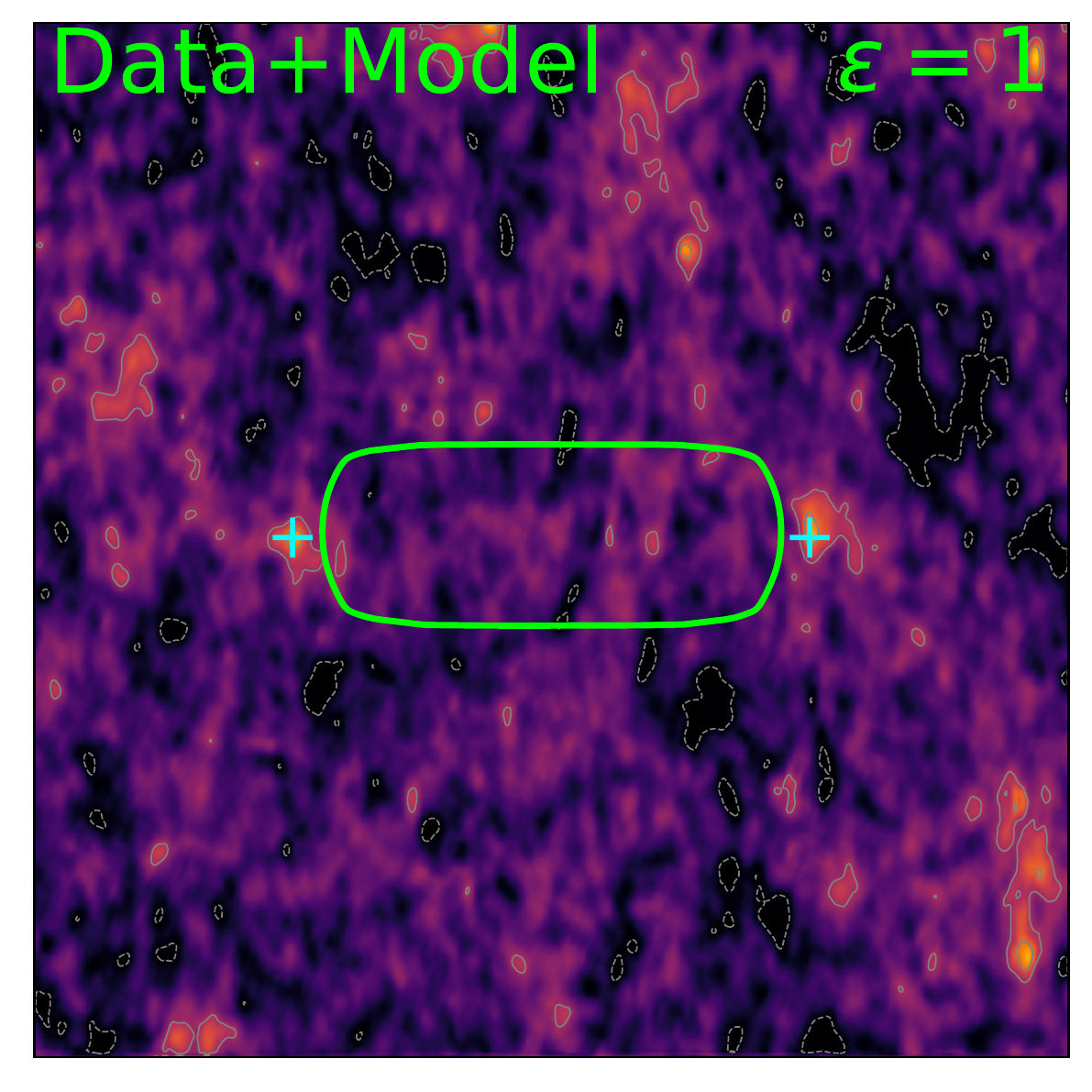}  \hfil
    \includegraphics[width=0.19\textwidth]{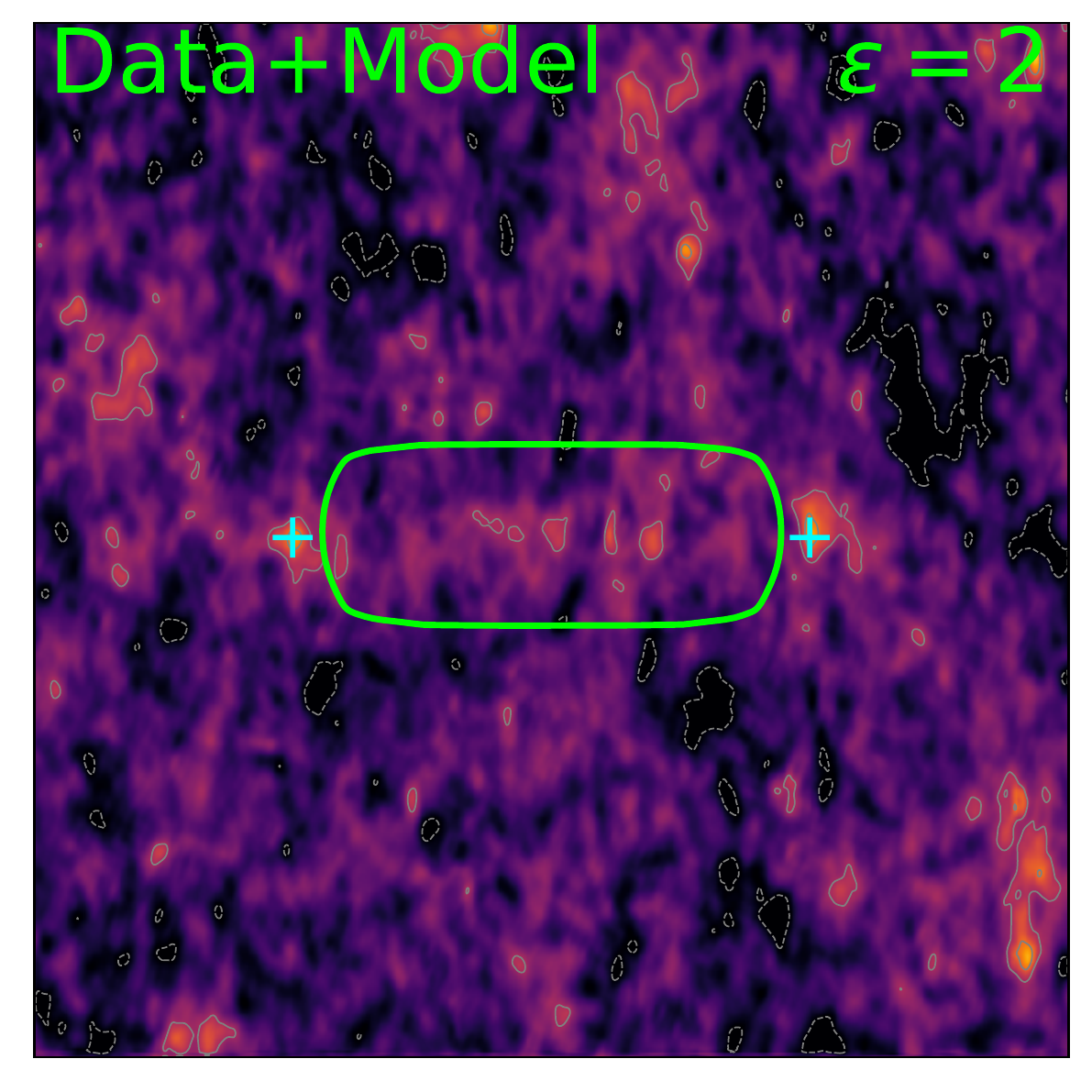}  \hfil
    \includegraphics[width=0.19\textwidth]{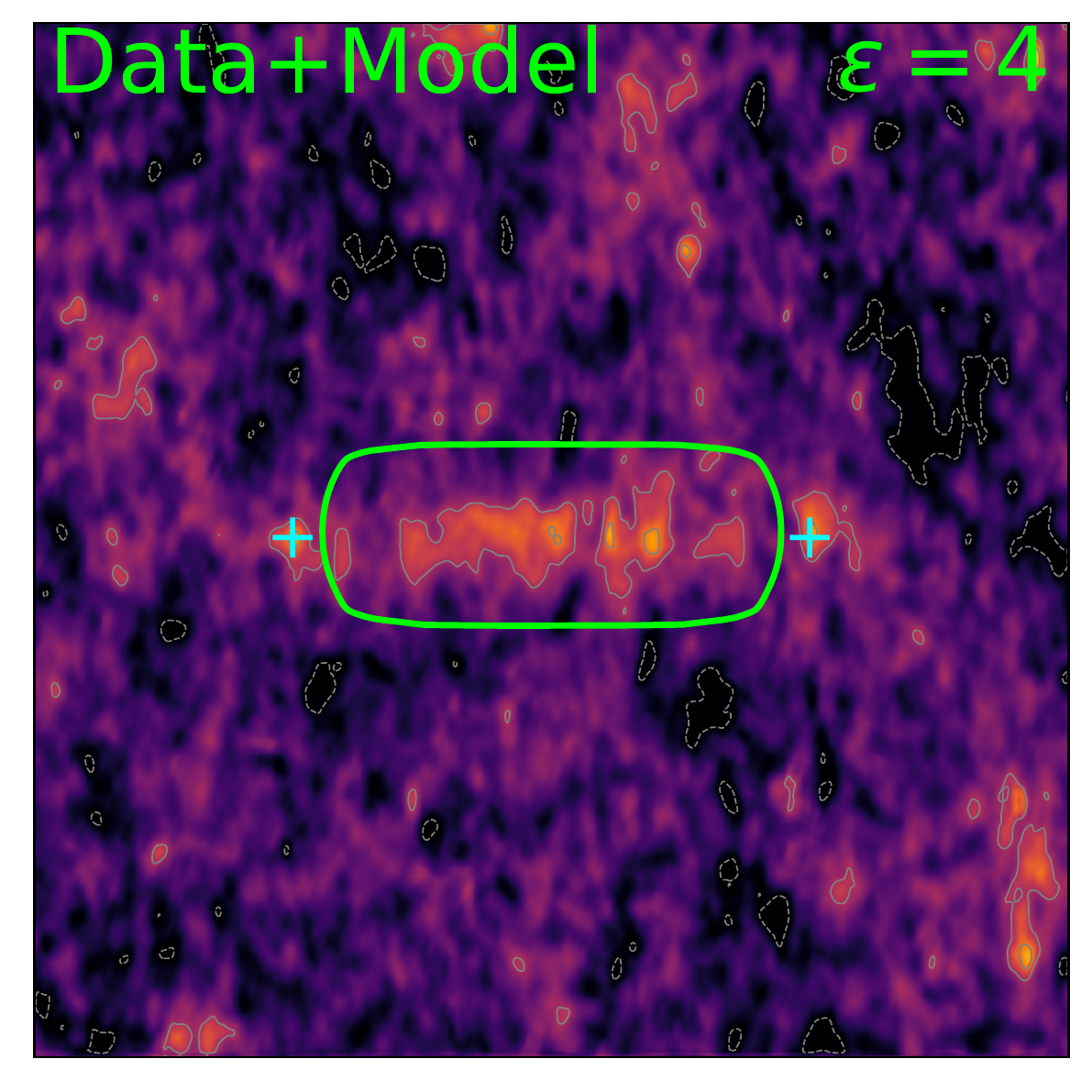}  \hfil
    \includegraphics[width=0.19\textwidth]{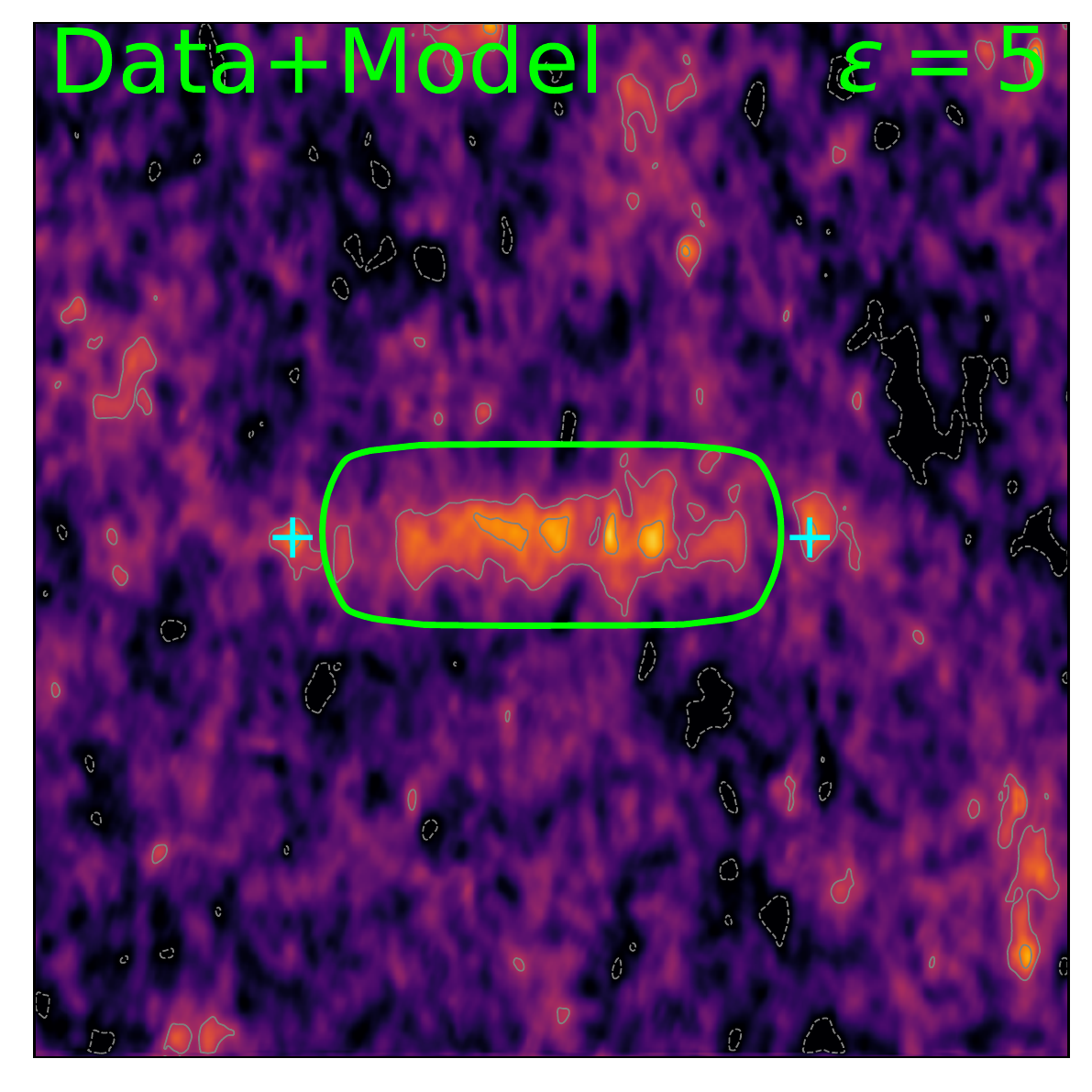}   };
\draw (-8.2, -1.27) node {\color{green} (\textit{a})};
\draw (-4.7, -1.27) node {\color{green} (\textit{b})};
\draw (-1.25, -1.27) node {\color{green} (\textit{c})};
\draw (2.15, -1.27) node {\color{green} (\textit{d})};
\draw (5.55, -1.27) node {\color{green} (\textit{e})};
\draw (-8.2, -4.75) node {\color{green} (\textit{f})};
\draw (-4.7, -4.75) node {\color{green} (\textit{g})};
\draw (-1.25, -4.75) node {\color{green} (\textit{h})};
\draw (2.15, -4.75) node {\color{green} (\textit{i})};
\draw (5.55, -4.75) node {\color{green} (\textit{j})};
\end{tikzpicture}
\caption{Stacked images of the cluster pairs before (\textit{a}) and after (\textit{b-e}, \textit{g-j}) injecting mock filaments (f) into \textit{uv} data (\textit{b-e}) and image (\textit{g-j}) with a range of emissivity (shown on the top right corners, in unit of $10^{-44} \, {\rm erg \, s^{-1} \, cm^{-3} \, Hz^{-1}}$). The original and model stacked images are shown in the panels (\textit{a}) and (\textit{f}). The green region is where the flux density is integrated. The gray contours start at $2\sigma_{\rm noise}$ and are spaced by a factor of 2. The negative contours are shown with dotted lines.}
\label{fig:stacked_inject}
\end{figure*}

The upper limit for the flux density of the filaments derived in the previous section is likely to be much higher than the true value that results in an unrealistic volume emissivity for the filaments.
To have a tight estimate of the emissivity, we perform the injection of the filament models in to the \textit{uv} data. More detail on the \textit{uv} injection procedure is described in Sec.~\ref{sec:model_inj}. To this end, we simulate the filaments as tube-like structures. We generate filament models with a range of emissivities $\epsilon_{\rm inj}=1-5\times10^{-44}\,{\rm erg \, s^{-1} \, cm^{-3} \, Hz^{-1}}$ (see Table~\ref{tab:stat}). For each injection, we assume a common emissivity for all the filaments in our sample. We create source-subtracted mosaics containing the filament models that are used to generate cutout images which are subsequently stacked, following the stacking technique described in Sec.~\ref{sec:stacking_radio}. In Fig.~\ref{fig:stacked_inject}, we present some of the stacked images before (\textit{a}) and after (\textit{b-e}) the \textit{uv} injection of the filament models. For comparison, we also show a stacked image of the filament models in Fig.~\ref{fig:stacked_inject} (\textit{f}). 

\begin{figure*}
\centering
\begin{tikzpicture}
\draw (0, 0) node[inner sep=0] {
    \includegraphics[width=0.33\textwidth]{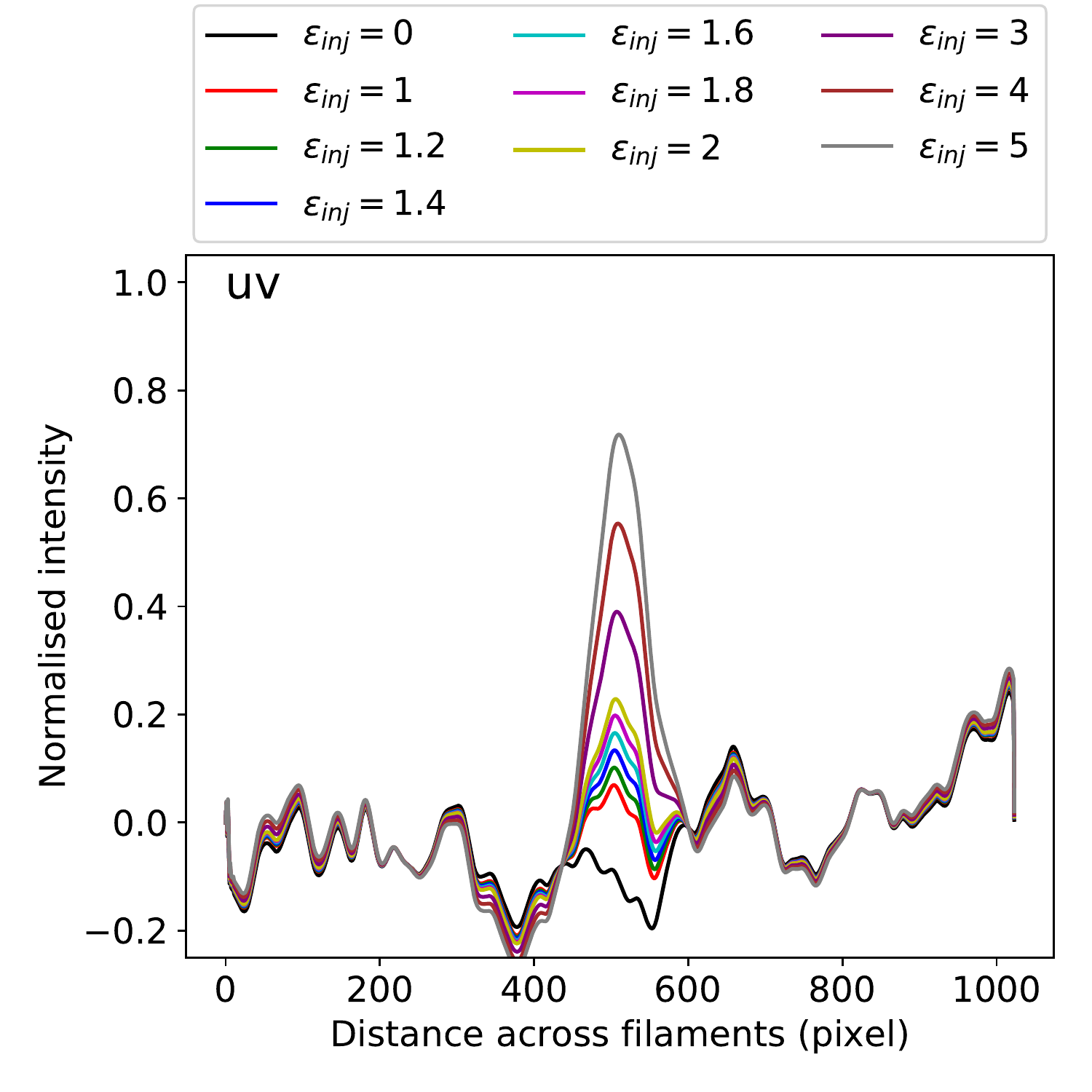}  \hfil
    \includegraphics[width=0.33\textwidth]{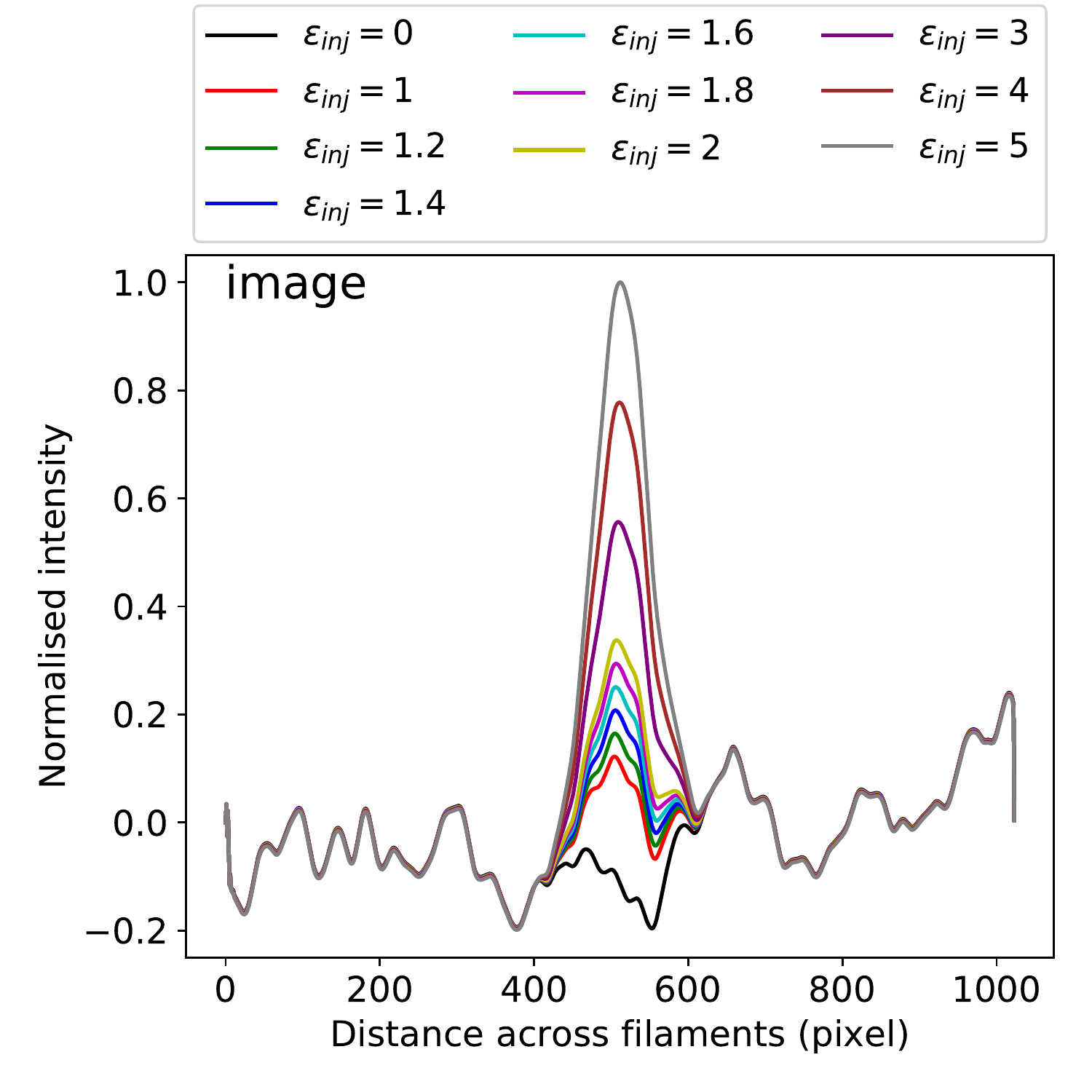}  \hfil
    \includegraphics[width=0.32\textwidth]{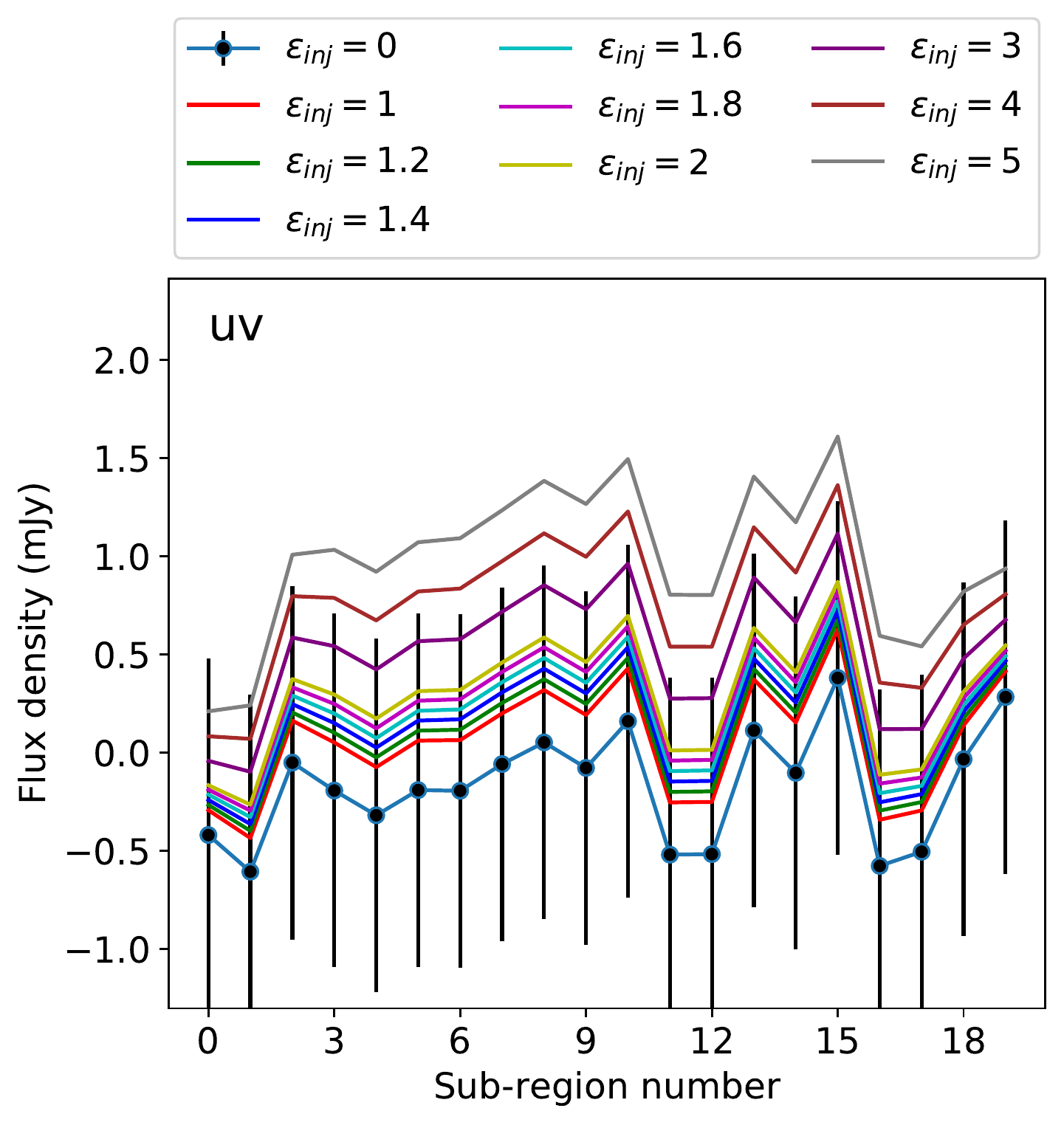}     };
\draw (-3.4, -2.1) node {\color{black} (\textit{a})};
\draw (2.475, -2.1) node {\color{black} (\textit{b})};
\draw (8.45, -2.12) node {\color{black} (\textit{c})};
\end{tikzpicture}
\caption{(a-b)Projected SB profiles across the widths of the stacked filaments with and without the injection of the models into \textit{uv} data (a) and images (b). The profiles (a,b) are normalised by dividing by the peak value. (c) The flux density extracted from the regions along the stacked filaments before and after the injection of filament models. The extracted region is shown with the white rectangle in Fig.~\ref{fig:stacked_inject} (\textit{e}). The injected emissivity in units of $10^{-44}{\rm erg \, s^{-1} \, cm^{-3} \, Hz^{-1}}$ is shown in the top  legends.
	}
	\label{fig:profs}
\end{figure*}

Fig.~\ref{fig:stacked_inject} (\textit{b-e}) shows that the SB in the stacked filament regions is brighter as the injected emissivity of the models increases. For the injected emissivity of $5\times10^{-44}\,{\rm erg \, s^{-1} \, cm^{-3} \, Hz^{-1}}$, diffuse emission is well detected at $2\sigma_{\rm noise}$ over the regions connecting the paired clusters (see Fig.~\ref{fig:stacked_inject}, \textit{e}). This feature is clearer in the normalised SB profiles across the widths of the stacked filaments in Fig.~\ref{fig:profs} (\textit{a}). As the injected emissivity increases, the peaks of the normalised intensity profiles increase. The intensity for $\epsilon_{\rm inj}=5\times10^{-44}\,{\rm erg \, s^{-1} \, cm^{-3} \, Hz^{-1}}$ is well above the background regions. We note that negative pixels are seen around the stacked filaments in Fig.~\ref{fig:stacked_inject} (\textit{b-e}) and the projected profiles in Fig.~\ref{fig:profs} (\textit{a}) (i.e. at the pixel locations of 370 and 610). These are due to the fact that the low SB of the injected emission is not deconvolved during the imaging (i.e. roundly below the noise levels). 

To search for the upper limit for the emissivity of the filaments, we compare the flux density extracted in the inter-cluster regions before and after the injection of filament models in to the \textit{uv} data. We select the region where the SB is brightest as shown by the white rectangle in Fig.~\ref{fig:stacked_inject} (\textit{e}). Following \cite{Vernstrom2021,Vernstrom2023}, we perform a comparison using two-sample Kolmogorov-Smirnov (KS) test\footnote{Available in \texttt{SciPy} package at \url{https://scipy.org}.} \citep{Hodges1958} that compares the two extracted flux density samples. One sample is taken from the injected stacked images (Fig.~\ref{fig:stacked_inject}, \textit{b-e}); and the other (reference) sample is extracted from the original stacked image (Fig.~\ref{fig:stacked_inject}, \textit{a}). The null hypothesis for the statistical test is that the flux density in the injected stacked image is not different from that in the original stacked image (i.e. the same distribution). We find that the injected emissivity of $1.2\times10^{-44} {\rm erg \, s^{-1} \, cm^{-3} \, Hz^{-1}}$ results in the statistically significant ($p$-value<0.05) difference in the flux densities in the injection and original stacked images. In addition to the comparison with the filament regions prior to the injection, we use background regions as the reference sample to compare the flux density in the injected filaments with the noise. The background regions are shown with the cyan rectangles in Fig.~\ref{fig:stacked_inject} (\textit{e}). We find that the upper limit obtained this way is consistent with the results from the previous case in which the reference sample is taken from the inter-cluster regions prior to the injection. More detail on the KS-test statistics is shown in Table~\ref{tab:stat}.

\begin{table}
\centering
 \caption{The statistics for the two-sample KS test for the comparison between the means of the emission in the inter-cluster filaments before and after the injection of filament models.} 
 \label{tab:stat}
\begin{tabular}{ cccc } 
 \hline\hline
  $\epsilon_{\rm inj}$ &  Statistic & $p$-value & Detection \\ 
 $[10^{-44} \, {\rm erg \, s^{-1} \, cm^{-3} \, Hz^{-1}}]$ &    &   & \\ \hline 
   1  &  0.4  & 0.08 & N \\
   1.2  &  0.45  & 0.03 & Y \\
   1.4  &  0.5  & 0.01 & Y \\
   1.6  &  0.55  & 0.004 & Y \\
   1.8  &  0.55  & 0.004 & Y \\
   2  &  0.6  &  0.001 & Y \\
   3  &  0.75  & <0.001 & Y \\
   4  &  0.85  & <0.001 & Y \\
   5  &  0.9  & <0.001 & Y \\
\hline\hline
\end{tabular}
\end{table}

\subsubsection{Equipartition magnetic fields}
\label{sec:bfields}

Diffuse synchrotron emission was reported in the regions connecting the pairs of LRGs that are thought to connect galaxy clusters \citep{Vernstrom2021,Vernstrom2023}. This implies the presence of large-scale magnetic fields in the inter-cluster regions. To estimate the magnetic field strength, we follow the equipartition assumption that the total energy in the magnetic field and the relativistic particles (i.e. electrons and heavy particles) is minimum. According to \cite{Pacholczyk1970,Govoni2004}, the equipartition magnetic field strength in the minimum total energy condition, in which the energies in the magnetic field and relativistic particles are approximately equal, is given by
\begin{equation} 
\label{eq:B}
    B_{\rm eq} = \left (   6\pi (1+k) c_{12} K_{\rm bol} \Phi^{-1} \epsilon\right )^{\nicefrac{2}{7}},
\end{equation}
where $k$ is the ratio of heavy particle (i.e. protons) energy to that in the electrons, $c_{12}(\alpha,\nu_1,\nu_2)=c_1^{\nicefrac{1}{2}} c_2^{-1} \tilde{C}$ with $c_1$, $c_2$ and $\tilde{C}$ given in Eq.~3 and 17 of \cite{Govoni2004}, $\nu_1$ and $\nu_2$ are the frequency range in which the synchrotron luminosity is integrated,  $K_{\rm bol}(z)=(1+z)^{(3+\alpha)}$ is the bolometric \textit{K}-correction term, $\Phi$ is the volume filling factor of the synchrotron emission, and $\epsilon$ is the volume emissivity of the synchrotron emission. In the equation above, we include the emissivity $\epsilon=\nicefrac{L_{\rm sync}}{V}$ and the K-correction term to the equation 22 of \cite{Govoni2004}.

As pointed out in literature  that the integration of the radio synchrotron luminosity should be taken in the electron energy space, instead of in the frequency one \citep{Brunetti2004}. This is because the radio emission from the relativistic electrons here is assumed to be in the frequency range from $\nu_1=10$~MHz and $\nu_2=10$~GHz (or 100~GHz) observable with radio telescopes on Earth which generates physical biases. The revised magnetic field strength is calculated based on the electron energy cutoff,
\begin{equation} 
\label{eq:B_rev}
    B^{'}_{\rm eq} \approx 1.1 \gamma_{\rm min} ^{\frac{1-2\alpha}{3+\alpha}} B_{\rm eq}^\frac{7}{2(3+\alpha)},
\end{equation}
where $\gamma_{\rm min}$ is the Lorentz factor of the minimum energy for the synchrotron radio emitting electrons \citep{Brunetti2004,Govoni2004}.

\begin{figure}
	\centering
	\includegraphics[width=1\columnwidth]{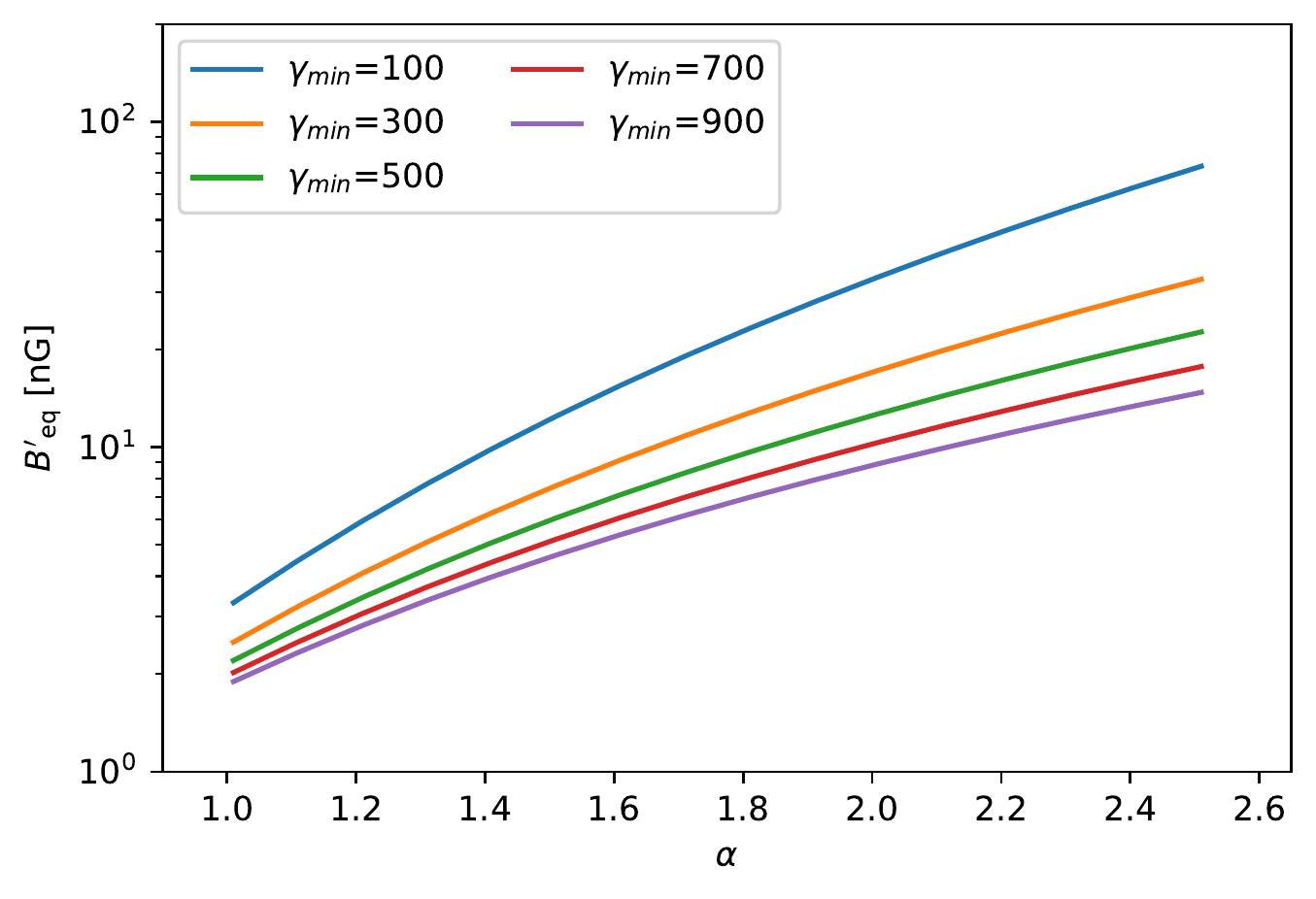} \\
	\caption{Magnetic field strength (revised) of cosmic filaments as a function of spectral index $\alpha$ and minimum particle energy cutoff $\gamma_{\rm min}$.
	}
	\label{fig:B}
\end{figure}

In the previous sections, we assume spectral index of 1.5 for the radio emission from the filaments although the true value for the spectral index is unknown. Here we examine the strength of magnetic field as a function of spectral index ranging between between 1 and 2.5. We assume the energy in the protons is 100 times that in relativistic electrons, i.e. $k=100$, that the emitted frequency ranges between 10~MHz and 10~GHz, and  $\Phi=1$ \citep[e.g.][]{Govoni2004}. In Fig.~\ref{fig:B} we present the dependence of the B field strength on the spectral index $\alpha$ and the minimum energy cutoff $\gamma_{\rm min}$ (assuming between 100 and 900). For our upper limit for the emissivity of $\epsilon=1.2\times10^{-44} {\rm erg \, s^{-1} \, cm^{-3} \, Hz^{-1}}$, the mean of revised magnetic field strength is well below $75\,{\rm nG}$ (i.e. $\gamma_{\rm min}=100$ and $\alpha=2.5$). Our constraint on the filament magnetic field is consistent with the estimates in the literature. For instance, \citealt{Vernstrom2021} report a magnetic field strength of $30-50 \, {\rm nG}$ using analysis of synchrotron -- Inverse-Compton emission. \cite{Carretti2022} estimate a strength of $B=32\pm3\,{\rm nG}$ using Rotation Measure technique. Our upper limit for the magnetic field strength is lower than those reported in \cite{OSullivan2019,Locatelli2021} (250~nG), \cite{Vernstrom2017,Brown2017} (30--200~nG) and is slightly higher than those in \cite{Vacca2018,Vernstrom2019,Amaral2021} (40--50~nG). We note that the true magnetic field strength in our calculation can be much lower in the cases that the spectral index is flatter and/or the minimum energy cutoff is higher. \cite{Vernstrom2021,Vernstrom2023} reported a mean spectral index of $\alpha\sim1$ for larger samples of (390,808 and 612,025, respectively) LRG pairs. If the magnetic field strength in the filaments is at the levels of that between the LRG pairs, our data suggests an upper limit of a few nG for the B field strength, as seen in Fig~\ref{fig:B}. However, this might not be necessary the case because of the differences (e.g. physical sizes, redshift, mass) in the samples used in this and \cite{Vernstrom2021}'s studies. For instance, the physical separation between the LRG pairs is below 14~Mpc (mean of 10~Mpc) that is much shorter than that of below 170~Mpc (mean of 40~Mpc) in our sample. Moreover, it is still unclear whether or not the magnetic field properties in the regions between the LRG pairs are similar to those in between the members of super-clusters used in this study.

\subsection{X-ray stacked profile}
\label{sec:res_xray}

\begin{figure}
	\centering
	\includegraphics[width=1\columnwidth]{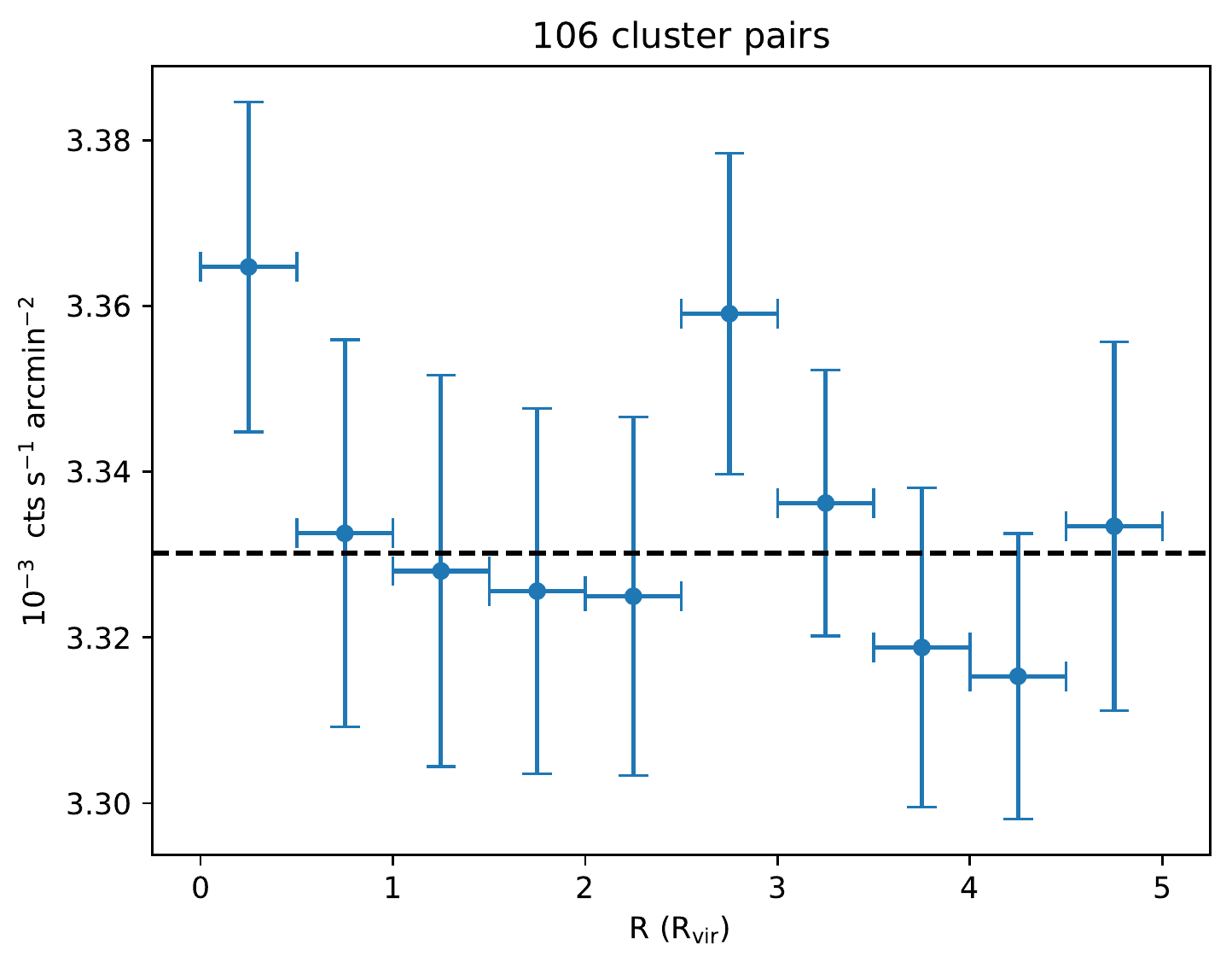} \\
	\caption{X-ray stacked profile. The dashed horizontal line denotes the mean surface brightness from 1 to 5 $\bar{R}_\mathrm{vir}$.
	}
	\label{fig:xray_stack}
\end{figure}

The mean and bootstrapping uncertainty of the radial profiles from 106 cluster pairs is plotted in Fig. \ref{fig:xray_stack}. The dashed line is the mean value from 1--5 $\bar{R}_\mathrm{vir}$, which is used as the local background to evaluate the excess emission within $\bar{R}_\mathrm{vir}$. Though the surface brightness in the first bin is marginally higher than the local background, the significance level of that bin is only 1.3 $\sigma$. Therefore, we do not detect significant excess of X-ray emission from the bridges of 106 cluster pairs.

\section{Discussion} 
\label{sec:disc}

\subsection{Radio emission}
\label{sec:dis_radio}

\begin{figure}
\centering
\includegraphics[width=1\columnwidth]{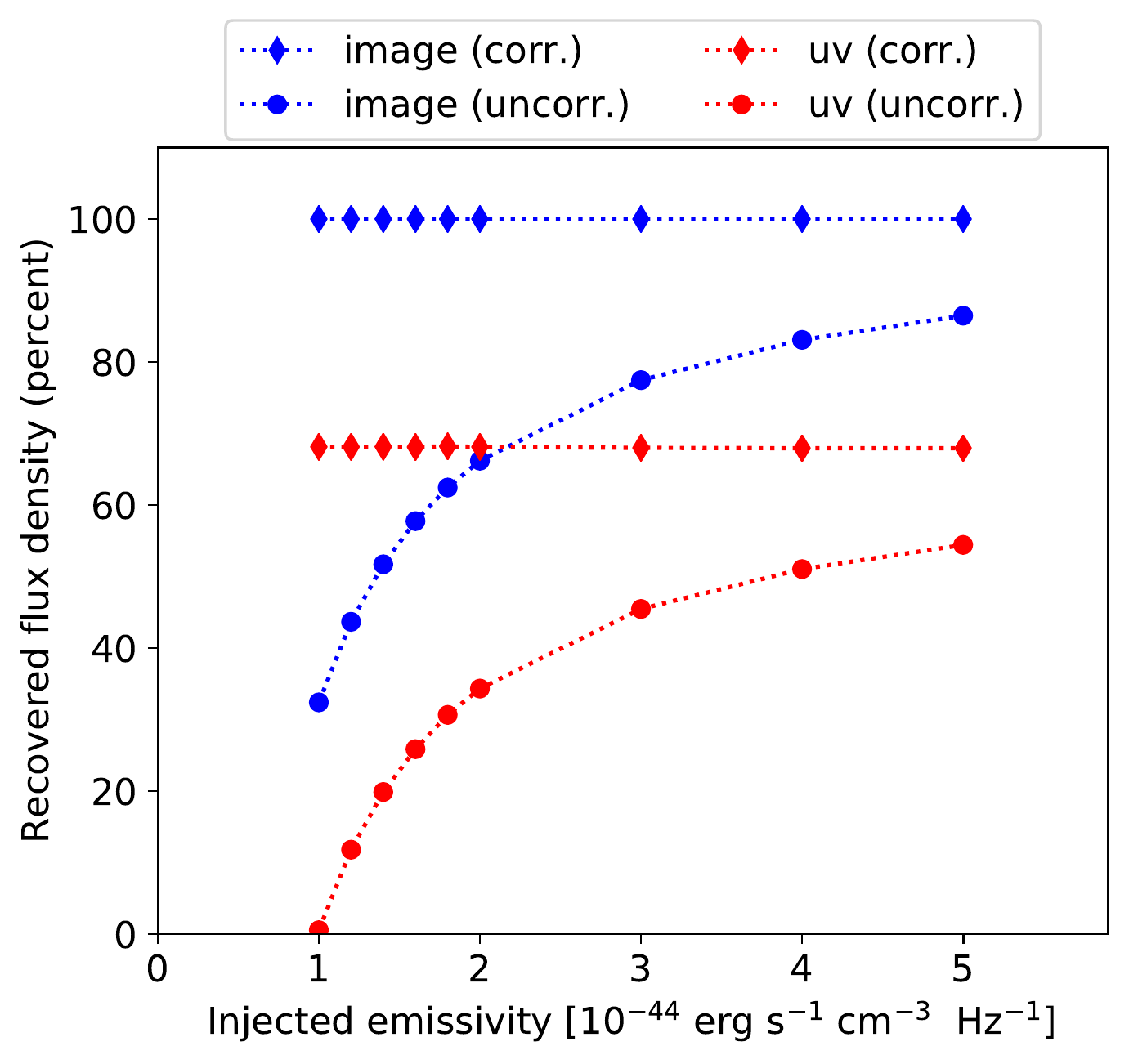} \\
\caption{The flux density is recovered after the injection of the models. The diamonds and circles indicate the recovered flux density with and without the subtraction of the pre-injection flux density, respectively. 
}
\label{fig:recover}
\end{figure}

In Fig.~\ref{fig:recover}, we present the dependence of the recovered flux density of the filament emission on the injected emissivity. The flux density is directly measured from the stacked regions between the cluster pairs before and after the \textit{uv} injection. We find that the fraction of the flux density detected with the LOFAR observations increases with the emissivity that is injected into the \textit{uv} data (i.e. red circles in Fig.~\ref{fig:recover}). It reaches to a value of $\sim$55 percent for the injected models of an emissivity of $\epsilon_{\rm inj}=5\times10^{-44}\,{\rm erg \, s^{-1} \, cm^{-3} \, Hz^{-1}}$. However, the trend could be biased by the non-zero emission in the inter-cluster regions prior to the injection. To correct for this effect, we subtract the flux density measured in the injected stacked image by the value measured in the original stacked image (i.e. without injection). After the correction, the fraction of the recovered flux density remains almost constant at $\sim$70 percent and does not depend on the injected emissivity. 

Possible reasons for the missing flux density (30 percent) could be due to (\textit{i}) the post-processing of the images caused by the stacking procedure 
and/or (\textit{ii}) the incomplete sampling of the \textit{uv} coverage. 
To examine the first possibility (\textit{i}), we perform the \textit{image} injection of the filament models into the cutout images as described in Sec.~\ref{sec:model_inj}. We then process the cutout images following the stacking procedure (i.e. rotating, regridding, and averaging). Here the models with a range of emissivity are created in a similar manner as for the \textit{uv} injection. The resulting stacked images and the flux density recovered after the injection into the cutout images are shown in Fig.~\ref{fig:stacked_inject} (\textit{g-j}) and Fig.~\ref{fig:recover}, respectively. We find that the uncorrected and corrected flux density through the injection into the cutout images follows similar patterns as those in the \textit{uv} injection. After the non-zero emission correction, the flux density measured in the stacked images is almost the same (i.e. 100 percent) as that in the stacked models (i.e. no flux density lost). Hence, the stacking procedure does not affect the missing flux density of the stacked filaments. Hence, the missing flux density after the \textit{uv} injection is likely due to the incomplete \textit{uv} coverage (\textit{ii}) of our LOFAR observations despite of the selection of the \textit{uv} range above $45\lambda$ that is sensitive to large-scale emission up to the scales of the filaments (i.e. $1.55\circ$).

Applying the injection procedure, we constrain the mean emissivity of the filaments to be lower than $1.2\times10^{-44} {\rm erg \, s^{-1} \, cm^{-3} \, Hz^{-1}}$. Our constraint is a factor of four higher than the estimate of $3.2\times10^{-45} {\rm erg \, s^{-1} \, cm^{-3} \, Hz^{-1}}$ that is reported for a sample of 390,808 stacked LRG pairs from the MWA and LWA by \cite{Vernstrom2021}. In addition, our constraint on the cosmic filaments is lower than the estimates for the inter-cluster regions for closer pairs of galaxy clusters reported in literature. For example, \cite{Govoni2019} determined an emissivity of $8.6 \times 10^{-43} {\rm erg \, s^{-1} \, cm^{-3} \, Hz^{-1}}$ at 140~MHz in the paired clusters Abell~399--Abell~401. \cite{Botteon2020} estimated a 140~MHz emissivity of $4 \times 10^{-43} {\rm erg \, s^{-1} \, cm^{-3} \, Hz^{-1}}$ for the radio diffuse emission connecting the pair Abell~1758N--Abell~1758S. The emissivity of the close cluster pairs is expected to be higher than that in the cosmic filaments. This might be the results of the non-linear amplification of the magnetic field strength and the turbulent re-acceleration of relativistic electrons via second-order Fermi mechanisms in the interacting regions of close pairs of galaxy clusters  \citep{Brunetti2020}.

According to the current models for the formation of magnetic field in cosmic filaments, the $B$ field can (\textit{i}) solely grow from a primordial field though magnetohydrodynamic (MHD) processes during the formation of large-scale structure \citep[e.g.][]{Vazza2015a} and/or (\textit{ii}) is seeding by astronomical sources \citep[e.g.][]{Aramburo-Garcia2021}. 
We constraint the upper limit for the magnetic field strength in the cosmic filaments to be between a few nG and 75~nG, depending on the assumed parameters such as the spectral index $\alpha$, the minimum energy $\gamma_{\rm min}$ of the radio emitting relativistic electrons, and the ratio $k$ between the heavy particle and electron energies. Due to the non-detection we are unable to estimate the spectral index for our sample. If the spectral index is similar to the value of $\sim1$ reported in \cite{Vernstrom2021,Vernstrom2023}, our data suggests that the $B$ field strength in the condition of minimum energy is not stronger than a few nG and weakly depends on $\gamma_{\rm min}$ (see Fig.~\ref{fig:B}). 
This is in line with the models in which the primordial magnetic field is amplified by MHD processes \citep[e.g.][]{Vazza2015a}. However, if the filaments have steep spectrum (e.g. $\alpha>2$), the upper limit for the $B$ field strength can be as high as several tens of nG. In this case, the models (\textit{i}) and (\textit{ii}) cannot be distinguished with our current data. Deeper radio observations will be required to  constrain the formation models of the $B$ field in cosmic filaments.

With the LOFAR data used in this work, the stacked image does not reach the sensitivity required for the detection of the radio diffuse emission from a small sample of 106 inter-cluster filaments. To detect the diffuse emission at the emissivity of $3.2\times10^{-45} {\rm erg \, s^{-1} \, cm^{-3} \, Hz^{-1}}$ that was reported by \cite{Vernstrom2021}, a stacked image with higher sensitivity of at least four times the current stacked image is required. This can be done by increasing the number of cluster pairs in the sample. To find the required number of the cluster pairs for LOFAR observations of equivalent sensitivity (i.e. 3.5~mJy~beam$^{-1}_{2\arcmin\times2\arcmin}$), we assume that the noise level in the stacked image decreases as the square root of the number of cluster pairs. Then, a sample of more than 1~700 cluster pairs is needed for the filament search through the \textit{uv} injection (see Sec.~\ref{sec:emissivity}). Moreover, to directly detect the filament emission from the stacked image at $2\sigma$, the noise needs to decrease by a factor of 8. Hence, a sample of at least 6~800 pairs of clusters is necessary for the detection which is a factor of 60 times less than the number of LRG pairs needed for the MWA and LWA in \cite{Vernstrom2021}. However, we note that the assumption on the dependence of the noise on the number of cutout images only holds when the paired clusters are separated by the same angular distance. Hence, the cutout images are not multiplied with a scaling factor during the re-scaling of the pixels when stacking.
In addition, the eFEDS field is located in an equatorial region (i.e. Dec is between $-2$ and $+5$) where LOFAR has low sensitivity. The typical noise RMS of the eFEDS mosaic is a factor of $2-3$ higher than that in a field of higher declination \citep[e.g.][]{Pasini2021,Shimwell2022}. Hence, the number of required pairs of clusters can be lower by a factor of $4-9$ times if the filaments are located at higher declinations. 

\subsection{X-ray emission}
\label{sec:dis_xray}
The X-ray emission in the 0.6--2.3 keV band traces the overdensity of the hot gas ($\sim10^7$~K) in the inter-cluster space. The major reason that we do not detect X-ray emission from the cluster pairs could be that we only have a limited sample of 106 pairs of clusters. 
With this sample size, we only have a noise level of $2\times10^{-5}$ cts s$^{-1}$ arcmin$^{-2}$ for a $0.5\times \bar{R}_{\mathrm{vir}}$ width bin. The statistics from this number is not high enough for detecting X-ray excess emission. As a comparison, the imaging analysis of \citet{Tanimura2022} includes 463 filaments and merely obtained $2.8\sigma$ significance, where the excess from the filaments center is $\sim$$4\times10^{-5}$ cts s$^{-1}$ arcmin$^{-2}$. 
Moreover, it is not clear whether the distribution of WHIM in the inter-cluster space follows the connecting axis of cluster pairs, especially for pairs with distances larger than 10~Mpc. 
The galaxy distribution from optical spectroscopic surveys could be a better proxy of the WHIM distribution in the cosmic web. For this reason, the stacked profile could be broaden and the detection significance could be decreased. Future study of filament stacking using the eRASS data will shed light on whether the optical defined filaments \citep[e.g.][]{Malavasi2020} or X-ray cluster pairs in superclusters defined using friend-to-friend algorithm \citep[e.g.][]{Liu2022} are the better proxy of the diffuse gas distribution in the inter-cluster space. 

\section{Conclusions}
\label{sec:con}

In this paper we present a search for inter-cluster filaments between members of superclusters of galaxies in the eFEDS field with the LOFAR and eROSITA observations. The observations do not detect the presence of diffuse X-ray and radio emission from the inter-cluster regions. The non-detection is most likely due to the small number (106) of cluster pairs in our sample and/or the limited sensitivity of observations. Using the radio data, we find that the mean radio emissivity of the filaments is below $1.2 \times 10^{-44} {\rm erg \, s^{-1} \, cm^{-3} \, Hz^{-1}}$. The mean strength for the magnetic field in the inter-cluster regions is weaker than 75~nG that depends on the spectral index of the radio emission and the minimum energy cutoff of the radio-emitting relativistic electrons. A tighter constraint on the magnetic field strength requires an estimate of the spectral index of the radio emission in filaments that provides more information on the origin of magnetic field in filaments (i.e. primordial or astronomical discrete sources). Future study with stacking of LOFAR and eROSITA survey data is likely to shed light on the detectabilities of the radio and X-ray emission from the inter-cluster filaments. 

\section*{Acknowledgements}
DNH and AB acknowledge support from the ERC through the grant ERC-StG DRANOEL n. 714245. 
MB acknowledges funding by the Deutsche Forschungsgemeinschaft (DFG, German Research Foundation) under Germany’s Excellence Strategy – EXC 2121 ‘Quantum Universe’ – 390833306.      
AB acknowledges support from the VIDI research programme with project number 639.042.729, which is financed by the Netherlands Organisation for Scientific Research (NWO). 
GDG acknowledges support from the Alexander von Humboldt Foundation. 
RJvW acknowledges support from the ERC Starting Grant ClusterWeb 804208. 
LOFAR (van Haarlem et al. 2013) is the Low Frequency Array designed and constructed by ASTRON. It has observing, data processing, and data storage facilities in several countries, which are owned by various parties (each with their own funding sources), and that are collectively operated by the ILT foundation under a joint scientific policy. The ILT resources have benefited from the following recent major funding sources: CNRS-INSU, Observatoire de Paris and Universit\'e d'Orl\'eans, France; BMBF, MIWF-NRW, MPG, Germany; Science Foundation Ireland (SFI), Department of Business, Enterprise and Innovation (DBEI), Ireland; NWO, The Netherlands; The Science and Technology Facilities Council, UK; Ministry of Science and Higher Education, Poland; The Istituto Nazionale di Astrofisica (INAF), Italy.
This research made use of the Dutch national e-infrastructure with support of the SURF Cooperative (e-infra 180169) and the LOFAR e-infra group. The J\"ulich LOFAR Long Term Archive and the German LOFAR network are both coordinated and operated by the J\"ulich Supercomputing Centre (JSC), and computing resources on the supercomputer JUWELS at JSC were provided by the Gauss Centre for Supercomputing e.V. (grant CHTB00) through the John von Neumann Institute for Computing (NIC).
This research made use of the University of Hertfordshire high-performance computing facility and the LOFAR-UK computing facility located at the University of Hertfordshire and supported by STFC [ST/P000096/1], and of the Italian LOFAR IT computing infrastructure supported and operated by INAF, and by the Physics Department of Turin university (under an agreement with Consorzio Interuniversitario per la Fisica Spaziale) at the C3S Supercomputing Centre, Italy.
This work is based on data from eROSITA, the soft X-ray instrument aboard SRG, a joint Russian-German science mission supported by the Russian Space Agency (Roskosmos), in the interests of the Russian Academy of Sciences represented by its Space Research Institute (IKI), and the Deutsches Zentrum f\"ur Luft- und Raumfahrt (DLR). The SRG spacecraft was built by Lavochkin Association (NPOL) and its subcontractors, and is operated by NPOL with support from the Max Planck Institute for Extraterrestrial Physics (MPE). The development and construction of the eROSITA X-ray instrument was led by MPE, with contributions from the Dr. Karl Remeis Observatory Bamberg \& ECAP (FAU Erlangen-Nuernberg), the University of Hamburg Observatory, the Leibniz Institute for Astrophysics Potsdam (AIP), and the Institute for Astronomy and Astrophysics of the University of T\"ubingen, with the support of DLR and the Max Planck Society. The Argelander Institute for Astronomy of the University of Bonn and the Ludwig Maximilians Universit\"at Munich also participated in the science preparation for eROSITA. The eROSITA data shown here were processed using the eSASS/NRTA software system developed by the German eROSITA consortium.

\section*{Data Availability}
The data underlying this article will be shared on reasonable request to the corresponding author.

\section*{ORCID IDs}
D. N. Hoang \orcidlink{0000-0002-8286-646X} \href{https://orcid.org/0000-0002-8286-646X}{https://orcid.org/0000-0002-8286-646X} \\
M. Br\"uggen \orcidlink{0000-0002-3369-7735} \href{https://orcid.org/0000-0002-3369-7735}{https://orcid.org/0000-0002-3369-7735} \\
X. Zhang \orcidlink{0000-0001-6019-6373} \href{https://orcid.org/0000-0001-6019-6373}{https://orcid.org/0000-0001-6019-6373} \\
A. Bonafede \orcidlink{0000-0002-5068-4581} \href{https://orcid.org/0000-0002-5068-4581}{https://orcid.org/0000-0002-5068-4581} \\
A. Liu \orcidlink{0000-0003-3501-0359} \href{https://orcid.org/0000-0003-3501-0359}{https://orcid.org/0000-0003-3501-0359} \\
T. Liu \orcidlink{0000-0002-2941-6734} \href{https://orcid.org/0000-0002-2941-6734}{https://orcid.org/0000-0002-2941-6734} \\
T. W. Shimwell \orcidlink{0000-0001-5648-9069} \href{https://orcid.org/0000-0001-5648-9069}{https://orcid.org/0000-0001-5648-9069} \\
A. Botteon \orcidlink{0000-0002-9325-1567} \href{https://orcid.org/0000-0002-9325-1567}{https://orcid.org/0000-0002-9325-1567} \\
E. Bulbul \orcidlink{0000-0002-7619-5399} \href{https://orcid.org/0000-0002-7619-5399}{https://orcid.org/0000-0002-7619-5399} \\
G. Di Gennaro \orcidlink{0000-0002-8648-8507} \href{https://orcid.org/0000-0002-8648-8507}{https://orcid.org/0000-0002-8648-8507} \\
S. P. O'Sullivan \orcidlink{0000-0002-3968-3051} \href{https://orcid.org/0000-0002-3968-3051}{https://orcid.org/0000-0002-3968-3051} \\
T. Pasini \orcidlink{0000-0002-9711-5554} \href{https://orcid.org/0000-0002-9711-5554}{https://orcid.org/0000-0002-9711-5554} \\
H. J. A. R\"ottgering \orcidlink{0000-0001-8887-2257} \href{https://orcid.org/0000-0001-8887-2257}{https://orcid.org/0000-0001-8887-2257} \\
R. J. van Weeren \orcidlink{0000-0002-0587-1660} \href{https://orcid.org/0000-0002-0587-1660}{https://orcid.org/0000-0002-0587-1660} \\



\bibliographystyle{mnras}




\appendix

\section{The sample of cluster pairs}
\label{sec:app_sample}
\begin{figure}
\centering
    \includegraphics[width=0.49\textwidth]{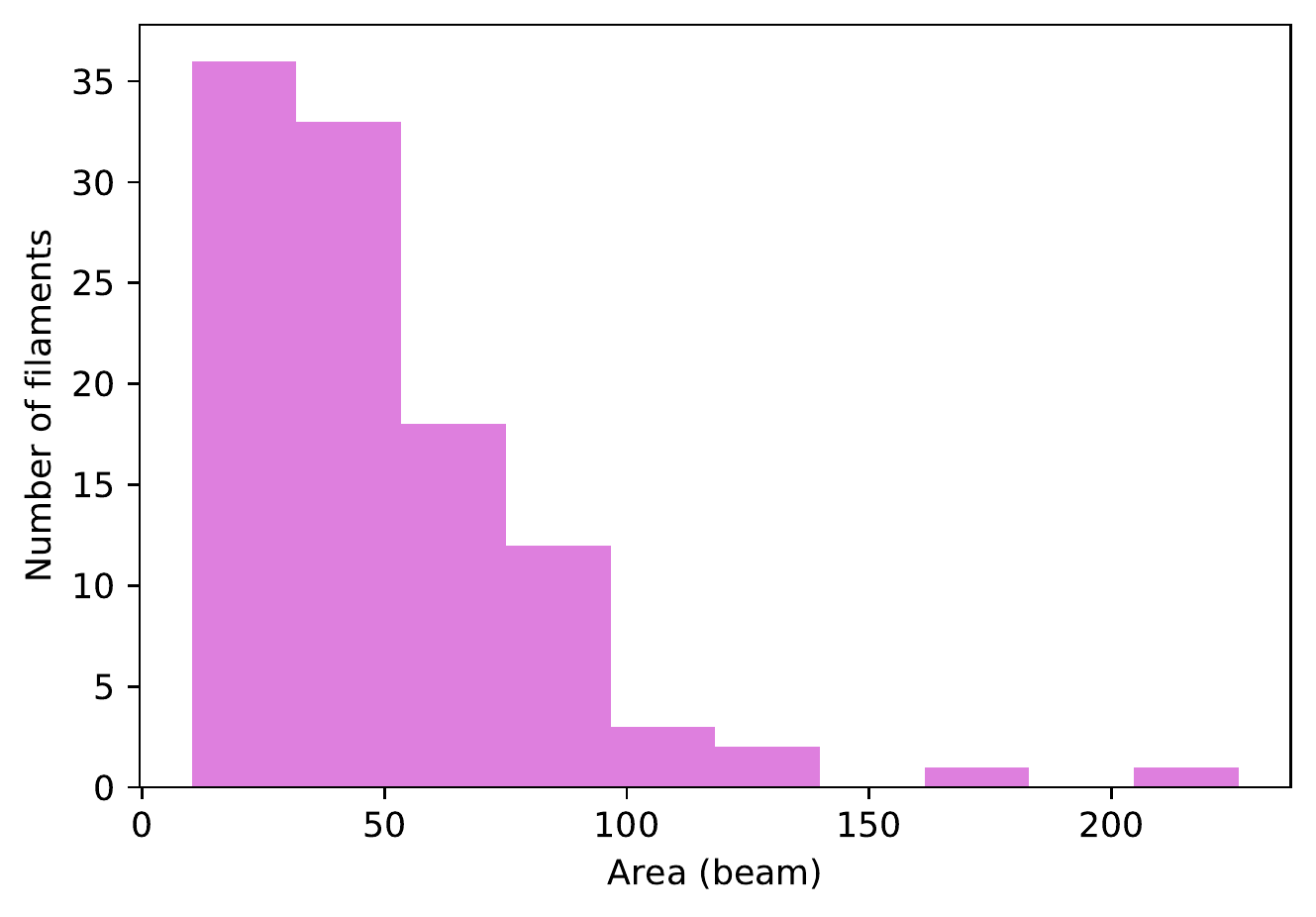} \hfil
    \includegraphics[width=0.49\textwidth]{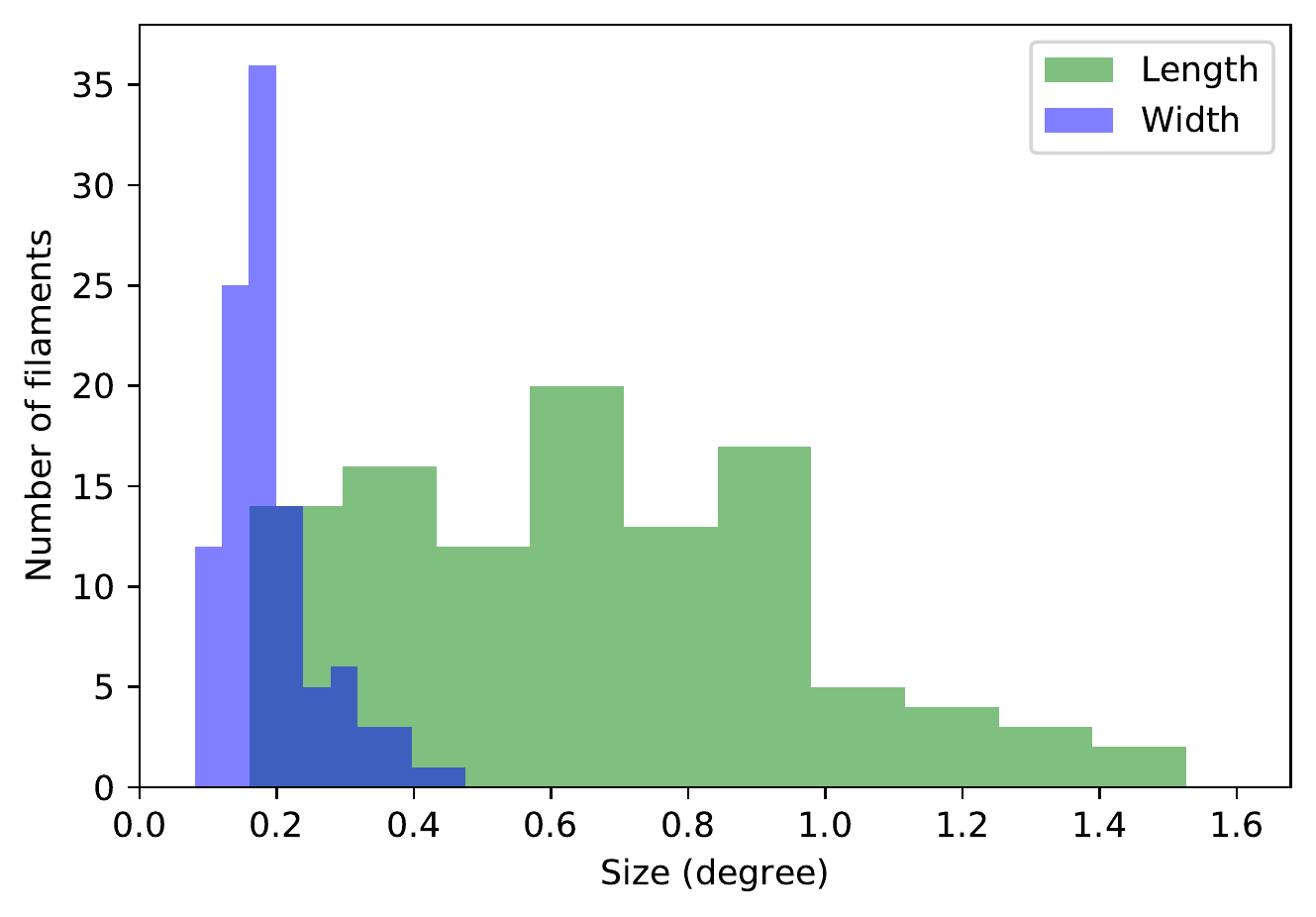}  \\
	\caption{Histogram of the area (\textit{top}) and angular sizes (\textit{bottom}) of the inter-cluster region.
	}
	\label{fig:fil_hist}
\end{figure}

%
In Fig.~\ref{fig:fil_hist} we show the distribution of the area for the cosmic filaments in our sample.

\onecolumn
\fontsize{8.0pt}{0cm}\selectfont

\begin{landscape}
\begin{longtable}{lllccccccccccccc}
\caption{\label{tab:pairs} The sample of inter-cluster filaments}\\
\hline\hline
ID & ID\_sc & Name & RA\_1 & Dec\_1 & z\_1 &  RA\_2 & Dec\_2 & z\_2 & RA\_mean & Dec\_mean & z\_mean  & Length & Width \\ \hline
\endfirsthead
\caption{continued.}\\
\hline\hline
ID & ID\_sc & Name & RA\_1 & Dec\_1 & z\_1 &  RA\_2 & Dec\_2 & z\_2 & RA\_mean & Dec\_mean & z\_mean &  Length & Width  \\ \hline
\hline
\endhead
\hline
\endfoot
1 & 1 & eFEDS J084034.5+023638--eFEDS J084430.8+021736 & 130.1441 & 2.6108 & 0.049 &  131.1285 & 2.2935 & 0.050 &  130.6363 & 2.4521 & 0.050 & 5.92 & 1.66 \\
2 & 1 & eFEDS J084430.8+021736--eFEDS J084645.6+014947 & 131.1285 & 2.2935 & 0.050 &  131.6902 & 1.8298 & 0.051 &  131.4094 & 2.0616 & 0.051 & 2.34 & 1.38 \\
3 & 2 & eFEDS J084531.6+022831--eFEDS J084148.1+004911 & 131.3818 & 2.4753 & 0.076 &  130.4507 & 0.8198 & 0.078 &  130.9163 & 1.6476 & 0.077 & 12.71 & 1.97 \\
4 & 2 & eFEDS J084148.1+004911--eFEDS J083807.6+002501 & 130.4507 & 0.8198 & 0.078 &  129.5321 & 0.4171 & 0.080 &  129.9914 & 0.6184 & 0.079 & 10.10 & 1.50 \\
5 & 3 & eFEDS J092629.3+032614--eFEDS J093056.9+034825 & 141.6225 & 3.4374 & 0.088 &  142.7372 & 3.8072 & 0.090 &  142.1799 & 3.6223 & 0.089 & 10.63 & 1.78 \\
6 & 3 & eFEDS J093056.9+034825--eFEDS J093253.8+025917 & 142.7372 & 3.8072 & 0.090 &  143.2243 & 2.9882 & 0.093 &  142.9808 & 3.3977 & 0.092 & 14.76 & 1.82 \\
7 & 4 & eFEDS J085436.6+003835--eFEDS J085705.9+011453 & 133.6526 & 0.6431 & 0.106 &  134.2749 & 1.2482 & 0.106 &  133.9637 & 0.9456 & 0.106 & 4.54 & 2.89 \\
8 & 4 & eFEDS J085141.9+021438--eFEDS J085705.9+011453 & 132.9248 & 2.2440 & 0.107 &  134.2749 & 1.2482 & 0.106 &  133.5999 & 1.7461 & 0.106 & 12.67 & 2.21 \\
9 & 4 & eFEDS J085433.0+004009--eFEDS J085705.9+011453 & 133.6376 & 0.6693 & 0.109 &  134.2749 & 1.2482 & 0.106 &  133.9563 & 0.9587 & 0.108 & 16.30 & 2.28 \\
10 & 5 & eFEDS J091336.6+031723--eFEDS J091453.6+041613 & 138.4026 & 3.2899 & 0.142 &  138.7235 & 4.2704 & 0.143 &  138.5630 & 3.7801 & 0.142 & 10.29 & 3.29 \\
11 & 6 & eFEDS J092955.8-003403--eFEDS J092953.5+002801 & 142.4829 & -0.5676 & 0.150 &  142.4733 & 0.4670 & 0.147 &  142.4781 & -0.0503 & 0.148 & 16.58 & 2.21 \\
12 & 6 & eFEDS J092953.5+002801--eFEDS J092735.3+014423 & 142.4733 & 0.4670 & 0.147 &  141.8974 & 1.7399 & 0.149 &  142.1853 & 1.1035 & 0.148 & 16.47 & 2.34 \\
13 & 8 & eFEDS J091412.6+001856--eFEDS J091403.3+013846 & 138.5528 & 0.3158 & 0.165 &  138.5140 & 1.6464 & 0.168 &  138.5334 & 0.9811 & 0.166 & 21.87 & 3.16 \\
14 & 10 & eFEDS J085438.5+001211--eFEDS J085508.9-003445 & 133.6608 & 0.2032 & 0.178 &  133.7874 & -0.5793 & 0.174 &  133.7241 & -0.1881 & 0.176 & 18.81 & 2.45 \\
15 & 12 & eFEDS J083651.3+030002--eFEDS J083955.0+022425 & 129.2138 & 3.0006 & 0.192 &  129.9793 & 2.4071 & 0.189 &  129.5966 & 2.7039 & 0.191 & 19.83 & 3.25 \\
16 & 13 & eFEDS J085030.5+003330--eFEDS J085327.2-002117 & 132.6272 & 0.5584 & 0.192 &  133.3634 & -0.3549 & 0.193 &  132.9953 & 0.1018 & 0.192 & 17.13 & 2.56 \\
17 & 13 & eFEDS J085327.2-002117--eFEDS J085022.2+001607 & 133.3634 & -0.3549 & 0.193 &  132.5927 & 0.2687 & 0.196 &  132.9780 & -0.0431 & 0.194 & 21.58 & 2.46 \\
18 & 13 & eFEDS J085340.5+022411--eFEDS J085128.4+011501 & 133.4191 & 2.4032 & 0.196 &  132.8685 & 1.2505 & 0.197 &  133.1438 & 1.8268 & 0.197 & 19.52 & 2.60 \\
19 & 13 & eFEDS J085027.8+001503--eFEDS J085128.4+011501 & 132.6160 & 0.2509 & 0.197 &  132.8685 & 1.2505 & 0.197 &  132.7423 & 0.7507 & 0.197 & 14.54 & 3.11 \\
20 & 14 & eFEDS J090137.7+030253--eFEDS J085913.1+031334 & 135.4072 & 3.0483 & 0.188 &  134.8048 & 3.2263 & 0.189 &  135.1060 & 3.1373 & 0.189 & 10.43 & 2.33 \\
21 & 14 & eFEDS J085913.1+031334--eFEDS J090119.0+030204 & 134.8048 & 3.2263 & 0.189 &  135.3294 & 3.0345 & 0.193 &  135.0671 & 3.1304 & 0.191 & 17.14 & 2.32 \\
22 & 14 & eFEDS J090131.1+030056--eFEDS J085931.9+030839 & 135.3800 & 3.0157 & 0.193 &  134.8830 & 3.1443 & 0.196 &  135.1315 & 3.0800 & 0.195 & 12.69 & 4.10 \\
23 & 14 & eFEDS J085931.9+030839--eFEDS J085728.3+032354 & 134.8830 & 3.1443 & 0.196 &  134.3680 & 3.3984 & 0.200 &  134.6255 & 3.2713 & 0.198 & 19.79 & 3.21 \\
24 & 14 & eFEDS J090255.2+030220--eFEDS J090010.4+023631 & 135.7300 & 3.0389 & 0.200 &  135.0435 & 2.6086 & 0.200 &  135.3867 & 2.8238 & 0.200 & 11.32 & 2.64 \\
25 & 14 & eFEDS J090010.4+023631--eFEDS J090200.5+022339 & 135.0435 & 2.6086 & 0.200 &  135.5024 & 2.3943 & 0.202 &  135.2729 & 2.5015 & 0.201 & 9.99 & 2.57 \\
26 & 14 & eFEDS J085751.6+031039--eFEDS J090200.5+022339 & 134.4653 & 3.1775 & 0.201 &  135.5024 & 2.3943 & 0.202 &  134.9839 & 2.7859 & 0.201 & 19.96 & 4.07 \\
27 & 15 & eFEDS J092241.9+020719--eFEDS J091858.0+024946 & 140.6749 & 2.1222 & 0.198 &  139.7418 & 2.8295 & 0.199 &  140.2084 & 2.4759 & 0.198 & 17.74 & 2.76 \\
28 & 16 & eFEDS J083204.4+041907--eFEDS J083412.7+035856 & 128.0185 & 4.3188 & 0.197 &  128.5529 & 3.9825 & 0.201 &  128.2857 & 4.1507 & 0.199 & 23.01 & 3.09 \\
29 & 17 & eFEDS J083431.0+034208--eFEDS J083817.4+041821 & 128.6296 & 3.7023 & 0.215 &  129.5727 & 4.3060 & 0.211 &  129.1011 & 4.0042 & 0.213 & 26.05 & 2.34 \\
30 & 19 & eFEDS J093012.0+030202--eFEDS J092647.5+030946 & 142.5503 & 3.0340 & 0.223 &  141.6981 & 3.1630 & 0.226 &  142.1242 & 3.0985 & 0.224 & 18.21 & 2.45 \\
31 & 21 & eFEDS J090723.8-011210--eFEDS J090843.9-013034 & 136.8494 & -1.2029 & 0.251 &  137.1831 & -1.5095 & 0.255 &  137.0163 & -1.3562 & 0.253 & 22.65 & 2.53 \\
32 & 22 & eFEDS J084751.7+025522--eFEDS J085051.8+015331 & 131.9657 & 2.9230 & 0.269 &  132.7161 & 1.8921 & 0.268 &  132.3409 & 2.4076 & 0.269 & 28.41 & 2.50 \\
33 & 23 & eFEDS J092022.8+045012--eFEDS J092246.2+034251 & 140.0954 & 4.8369 & 0.270 &  140.6928 & 3.7143 & 0.269 &  140.3941 & 4.2756 & 0.269 & 28.84 & 2.95 \\
34 & 24 & eFEDS J084151.9-010156--eFEDS J083921.0-014149 & 130.4665 & -1.0323 & 0.270 &  129.8377 & -1.6970 & 0.269 &  130.1521 & -1.3647 & 0.269 & 19.80 & 2.79 \\
35 & 26 & eFEDS J092740.1+042038--eFEDS J092928.3+042411 & 141.9173 & 4.3440 & 0.275 &  142.3679 & 4.4032 & 0.278 &  142.1426 & 4.3736 & 0.277 & 18.78 & 3.02 \\
36 & 28 & eFEDS J092031.3+024710--eFEDS J092053.4+021125 & 140.1306 & 2.7863 & 0.278 &  140.2228 & 2.1905 & 0.280 &  140.1767 & 2.4884 & 0.279 & 16.67 & 2.68 \\
37 & 28 & eFEDS J092053.4+021125--eFEDS J091851.7+021432 & 140.2228 & 2.1905 & 0.280 &  139.7155 & 2.2423 & 0.283 &  139.9692 & 2.2164 & 0.282 & 17.74 & 2.69 \\
38 & 28 & eFEDS J091849.0+021204--eFEDS J092049.5+024513 & 139.7042 & 2.2013 & 0.283 &  140.2063 & 2.7538 & 0.284 &  139.9553 & 2.4776 & 0.283 & 15.24 & 3.95 \\
39 & 30 & eFEDS J085335.2+032214--eFEDS J085020.4+032819 & 133.3967 & 3.3708 & 0.295 &  132.5851 & 3.4722 & 0.294 &  132.9909 & 3.4215 & 0.294 & 19.06 & 2.98 \\
40 & 31 & eFEDS J083137.9+004632--eFEDS J083503.2+010756 & 127.9081 & 0.7758 & 0.293 &  128.7636 & 1.1325 & 0.296 &  128.3359 & 0.9541 & 0.294 & 27.12 & 2.88 \\
41 & 32 & eFEDS J090053.0-002837--eFEDS J090153.9-012209 & 135.2212 & -0.4772 & 0.295 &  135.4747 & -1.3694 & 0.295 &  135.3480 & -0.9233 & 0.295 & 21.79 & 2.87 \\
42 & 33 & eFEDS J091543.5-004944--eFEDS J091351.1-004507 & 138.9316 & -0.8291 & 0.296 &  138.4632 & -0.7520 & 0.294 &  138.6974 & -0.7905 & 0.295 & 13.09 & 2.98 \\
43 & 34 & eFEDS J091248.2+002446--eFEDS J090913.8-001214 & 138.2011 & 0.4130 & 0.308 &  137.3078 & -0.2040 & 0.310 &  137.7545 & 0.1045 & 0.309 & 29.05 & 3.57 \\
44 & 35 & eFEDS J090115.3+005040--eFEDS J090409.7+003831 & 135.3140 & 0.8445 & 0.312 &  136.0404 & 0.6421 & 0.311 &  135.6772 & 0.7433 & 0.312 & 18.97 & 2.79 \\
45 & 35 & eFEDS J090703.9+010756--eFEDS J090409.7+003831 & 136.7663 & 1.1323 & 0.307 &  136.0404 & 0.6421 & 0.311 &  136.4034 & 0.8872 & 0.309 & 31.17 & 2.80 \\
46 & 38 & eFEDS J092813.0-004508--eFEDS J092915.7-001357 & 142.0544 & -0.7524 & 0.318 &  142.3158 & -0.2327 & 0.322 &  142.1851 & -0.4926 & 0.320 & 26.87 & 2.74 \\
47 & 39 & eFEDS J092212.0-002731--eFEDS J092312.0+000355 & 140.5503 & -0.4586 & 0.321 &  140.8003 & 0.0655 & 0.321 &  140.6753 & -0.1966 & 0.321 & 14.01 & 3.47 \\
48 & 40 & eFEDS J085517.2+013508--eFEDS J085121.2+012856 & 133.8219 & 1.5857 & 0.324 &  132.8384 & 1.4825 & 0.327 &  133.3302 & 1.5341 & 0.325 & 31.79 & 2.90 \\
49 & 40 & eFEDS J085624.3+004632--eFEDS J085121.2+012856 & 134.1016 & 0.7757 & 0.319 &  132.8384 & 1.4825 & 0.327 &  133.4700 & 1.1291 & 0.323 & 59.26 & 2.85 \\
50 & 40 & eFEDS J085121.2+012856--eFEDS J084934.9+014437 & 132.8384 & 1.4825 & 0.327 &  132.3958 & 1.7438 & 0.325 &  132.6171 & 1.6131 & 0.326 & 14.82 & 3.09 \\
51 & 42 & eFEDS J084135.7-005048--eFEDS J084324.2-001438 & 130.3988 & -0.8468 & 0.328 &  130.8511 & -0.2441 & 0.330 &  130.6250 & -0.5454 & 0.329 & 23.99 & 2.68 \\
52 & 43 & eFEDS J090817.2-013034--eFEDS J090916.0-015540 & 137.0717 & -1.5096 & 0.331 &  137.3168 & -1.9280 & 0.326 &  137.1942 & -1.7188 & 0.329 & 30.23 & 2.73 \\
53 & 44 & eFEDS J093151.3-002212--eFEDS J093049.2-003714 & 142.9638 & -0.3701 & 0.336 &  142.7052 & -0.6207 & 0.334 &  142.8345 & -0.4954 & 0.335 & 10.97 & 3.44 \\
54 & 45 & eFEDS J092739.7-010427--eFEDS J092740.7-015320 & 141.9158 & -1.0743 & 0.329 &  141.9196 & -1.8889 & 0.332 &  141.9177 & -1.4816 & 0.331 & 24.72 & 3.14 \\
55 & 45 & eFEDS J092740.7-015320--eFEDS J092548.9-011725 & 141.9196 & -1.8889 & 0.332 &  141.4539 & -1.2903 & 0.337 &  141.6867 & -1.5896 & 0.334 & 35.41 & 2.94 \\
56 & 45 & eFEDS J092548.9-011725--eFEDS J092405.0-013059 & 141.4539 & -1.2903 & 0.337 &  141.0211 & -1.5165 & 0.337 &  141.2375 & -1.4034 & 0.337 & 12.15 & 3.43 \\
57 & 45 & eFEDS J092405.0-013059--eFEDS J092621.3-003356 & 141.0211 & -1.5165 & 0.337 &  141.5890 & -0.5657 & 0.340 &  141.3050 & -1.0411 & 0.339 & 35.89 & 3.27 \\
58 & 45 & eFEDS J092621.3-003356--eFEDS J092846.5+000056 & 141.5890 & -0.5657 & 0.340 &  142.1940 & 0.0157 & 0.344 &  141.8915 & -0.2750 & 0.342 & 33.24 & 3.00 \\
59 & 46 & eFEDS J084253.7+002006--eFEDS J084346.2+010833 & 130.7238 & 0.3350 & 0.345 &  130.9425 & 1.1425 & 0.342 &  130.8331 & 0.7388 & 0.343 & 29.68 & 2.56 \\
60 & 46 & eFEDS J084346.2+010833--eFEDS J084004.8+013751 & 130.9425 & 1.1425 & 0.342 &  130.0203 & 1.6309 & 0.342 &  130.4814 & 1.3867 & 0.342 & 30.23 & 2.71 \\
61 & 47 & eFEDS J093431.3-002309--eFEDS J093546.3-000115 & 143.6304 & -0.3860 & 0.342 &  143.9433 & -0.0211 & 0.339 &  143.7868 & -0.2035 & 0.341 & 18.51 & 3.40 \\
62 & 47 & eFEDS J093316.6+004619--eFEDS J093544.2-000339 & 143.3195 & 0.7721 & 0.347 &  143.9342 & -0.0609 & 0.347 &  143.6269 & 0.3556 & 0.347 & 30.11 & 3.19 \\
63 & 48 & eFEDS J092246.4+042424--eFEDS J092644.0+040010 & 140.6936 & 4.4067 & 0.346 &  141.6834 & 4.0029 & 0.347 &  141.1885 & 4.2048 & 0.347 & 31.77 & 2.73 \\
64 & 49 & eFEDS J084558.1+012443--eFEDS J084729.7+013053 & 131.4924 & 1.4120 & 0.350 &  131.8740 & 1.5149 & 0.351 &  131.6832 & 1.4635 & 0.350 & 11.72 & 2.98 \\
65 & 50 & eFEDS J092744.6+045630--eFEDS J092918.3+044925 & 141.9360 & 4.9418 & 0.355 &  142.3266 & 4.8237 & 0.352 &  142.1313 & 4.8828 & 0.354 & 18.24 & 2.97 \\
66 & 51 & eFEDS J091749.4+014621--eFEDS J091722.4+010118 & 139.4559 & 1.7728 & 0.355 &  139.3435 & 1.0218 & 0.359 &  139.3997 & 1.3973 & 0.357 & 33.38 & 3.31 \\
67 & 52 & eFEDS J093500.7+005417--eFEDS J093612.7+001650 & 143.7532 & 0.9048 & 0.361 &  144.0529 & 0.2807 & 0.358 &  143.9031 & 0.5927 & 0.360 & 24.96 & 3.49 \\
68 & 52 & eFEDS J093612.7+001650--eFEDS J093513.0+004757 & 144.0529 & 0.2807 & 0.358 &  143.8046 & 0.7994 & 0.356 &  143.9287 & 0.5400 & 0.357 & 19.81 & 4.21 \\
69 & 53 & eFEDS J084021.6+020132--eFEDS J083900.6+020057 & 130.0903 & 2.0256 & 0.357 &  129.7527 & 2.0159 & 0.359 &  129.9215 & 2.0208 & 0.358 & 11.80 & 2.12 \\
70 & 54 & eFEDS J090224.4-005150--eFEDS J090105.2-012525 & 135.6020 & -0.8641 & 0.409 &  135.2718 & -1.4237 & 0.405 &  135.4369 & -1.1439 & 0.407 & 35.53 & 2.92 \\
71 & 55 & eFEDS J090059.3+035925--eFEDS J090323.7+030738 & 135.2471 & 3.9905 & 0.412 &  135.8489 & 3.1273 & 0.407 &  135.5480 & 3.5589 & 0.410 & 46.46 & 3.51 \\
72 & 56 & eFEDS J084051.7+014122--eFEDS J084110.8+005200 & 130.2156 & 1.6895 & 0.411 &  130.2953 & 0.8668 & 0.415 &  130.2554 & 1.2782 & 0.413 & 38.11 & 2.87 \\
73 & 56 & eFEDS J084110.8+005200--eFEDS J084220.9+013844 & 130.2953 & 0.8668 & 0.415 &  130.5875 & 1.6457 & 0.421 &  130.4414 & 1.2563 & 0.418 & 46.29 & 1.63 \\
74 & 56 & eFEDS J084649.0+004946--eFEDS J084501.0+012728 & 131.7045 & 0.8295 & 0.416 &  131.2542 & 1.4578 & 0.420 &  131.4794 & 1.1437 & 0.418 & 37.63 & 2.95 \\
75 & 56 & eFEDS J084501.0+012728--eFEDS J084220.9+013844 & 131.2542 & 1.4578 & 0.420 &  130.5875 & 1.6457 & 0.421 &  130.9209 & 1.5518 & 0.420 & 25.88 & 2.04 \\
76 & 56 & eFEDS J084220.9+013844--eFEDS J084210.5+020558 & 130.5875 & 1.6457 & 0.421 &  130.5439 & 2.0997 & 0.421 &  130.5657 & 1.8727 & 0.421 & 17.12 & 1.61 \\
77 & 56 & eFEDS J084210.5+020558--eFEDS J084129.0+002645 & 130.5439 & 2.0997 & 0.421 &  130.3708 & 0.4460 & 0.402 &  130.4574 & 1.2728 & 0.412 & 136.07 & 2.75 \\
78 & 58 & eFEDS J083228.0-000656--eFEDS J082952.7+002139 & 128.1169 & -0.1157 & 0.421 &  127.4697 & 0.3611 & 0.420 &  127.7933 & 0.1227 & 0.421 & 30.92 & 2.79 \\
79 & 59 & eFEDS J085950.1-001221--eFEDS J090044.6-011104 & 134.9590 & -0.2058 & 0.427 &  135.1862 & -1.1845 & 0.430 &  135.0726 & -0.6951 & 0.428 & 44.46 & 1.68 \\
80 & 61 & eFEDS J091302.1+035000--eFEDS J091522.5+041201 & 138.2590 & 3.8336 & 0.455 &  138.8438 & 4.2003 & 0.460 &  138.5514 & 4.0169 & 0.457 & 44.97 & 3.08 \\
81 & 62 & eFEDS J084006.1+025913--eFEDS J083834.1+020643 & 130.0258 & 2.9871 & 0.456 &  129.6425 & 2.1121 & 0.457 &  129.8341 & 2.5496 & 0.457 & 39.18 & 3.37 \\
82 & 63 & eFEDS J085217.0-010131--eFEDS J085138.2-003537 & 133.0709 & -1.0254 & 0.460 &  132.9095 & -0.5938 & 0.458 &  132.9902 & -0.8096 & 0.459 & 20.30 & 3.74 \\
83 & 64 & eFEDS J090627.5+035846--eFEDS J090430.7+042648 & 136.6148 & 3.9795 & 0.463 &  136.1282 & 4.4468 & 0.457 &  136.3715 & 4.2132 & 0.460 & 51.06 & 3.27 \\
84 & 65 & eFEDS J092022.0+030106--eFEDS J091741.1+024518 & 140.0918 & 3.0186 & 0.481 &  139.4215 & 2.7551 & 0.486 &  139.7567 & 2.8868 & 0.483 & 47.37 & 2.67 \\
85 & 66 & eFEDS J091555.6-013248--eFEDS J091439.5-014416 & 138.9819 & -1.5468 & 0.490 &  138.6648 & -1.7380 & 0.484 &  138.8234 & -1.6424 & 0.487 & 38.87 & 3.68 \\
86 & 67 & eFEDS J092828.3-000955--eFEDS J092900.0-003920 & 142.1180 & -0.1653 & 0.509 &  142.2502 & -0.6557 & 0.503 &  142.1841 & -0.4105 & 0.506 & 44.12 & 2.72 \\
87 & 69 & eFEDS J090210.6+032513--eFEDS J090417.0+040439 & 135.5445 & 3.4204 & 0.535 &  136.0712 & 4.0776 & 0.536 &  135.8079 & 3.7490 & 0.535 & 43.23 & 2.80 \\
88 & 71 & eFEDS J083811.8-015934--eFEDS J083427.0-015612 & 129.5496 & -1.9930 & 0.560 &  128.6129 & -1.9369 & 0.573 &  129.0812 & -1.9649 & 0.566 & 101.35 & 3.85 \\
89 & 71 & eFEDS J083427.0-015612--eFEDS J083929.6-015005 & 128.6129 & -1.9369 & 0.573 &  129.8736 & -1.8348 & 0.575 &  129.2432 & -1.8859 & 0.574 & 72.12 & 3.02 \\
90 & 73 & eFEDS J092041.1+041117--eFEDS J092258.2+032041 & 140.1716 & 4.1883 & 0.580 &  140.7426 & 3.3449 & 0.575 &  140.4571 & 3.7666 & 0.577 & 66.17 & 2.88 \\
91 & 74 & eFEDS J091139.3-014144--eFEDS J091300.9-013152 & 137.9141 & -1.6958 & 0.589 &  138.2538 & -1.5311 & 0.592 &  138.0840 & -1.6135 & 0.590 & 24.77 & 3.08 \\
92 & 75 & eFEDS J091254.4+032028--eFEDS J091648.1+030506 & 138.2270 & 3.3414 & 0.619 &  139.2007 & 3.0851 & 0.620 &  138.7138 & 3.2132 & 0.620 & 61.16 & 3.46 \\
93 & 75 & eFEDS J090336.7+033124--eFEDS J090751.9+024647 & 135.9033 & 3.5235 & 0.617 &  136.9666 & 2.7797 & 0.618 &  136.4350 & 3.1516 & 0.618 & 79.88 & 2.87 \\
94 & 76 & eFEDS J084939.6-005126--eFEDS J084833.2-012216 & 132.4151 & -0.8573 & 0.613 &  132.1385 & -1.3712 & 0.629 &  132.2768 & -1.1142 & 0.621 & 113.36 & 3.77 \\
95 & 77 & eFEDS J090750.1+025006--eFEDS J090540.7+013219 & 136.9591 & 2.8351 & 0.648 &  136.4199 & 1.5388 & 0.644 &  136.6895 & 2.1869 & 0.646 & 94.65 & 3.31 \\
96 & 77 & eFEDS J090805.9+011952--eFEDS J090540.7+013219 & 137.0250 & 1.3314 & 0.659 &  136.4199 & 1.5388 & 0.644 &  136.7224 & 1.4351 & 0.651 & 114.04 & 3.20 \\
97 & 78 & eFEDS J083403.7+012131--eFEDS J083120.5+005257 & 128.5155 & 1.3588 & 0.671 &  127.8354 & 0.8825 & 0.664 &  128.1755 & 1.1206 & 0.667 & 74.89 & 2.89 \\
98 & 78 & eFEDS J083125.9+015533--eFEDS J083120.5+005257 & 127.8580 & 1.9259 & 0.684 &  127.8354 & 0.8825 & 0.664 &  127.8467 & 1.4042 & 0.674 & 166.36 & 3.13 \\
99 & 79 & eFEDS J084434.3+031026--eFEDS J084717.7+033421 & 131.1430 & 3.1739 & 0.724 &  131.8241 & 3.5725 & 0.716 &  131.4835 & 3.3732 & 0.720 & 82.29 & 2.81 \\
100 & 80 & eFEDS J085627.2+014217--eFEDS J085837.9+012657 & 134.1134 & 1.7050 & 0.732 &  134.6581 & 1.4494 & 0.750 &  134.3857 & 1.5772 & 0.741 & 144.20 & 3.50 \\
101 & 81 & eFEDS J090806.4+032613--eFEDS J090930.6+034055 & 137.0270 & 3.4370 & 0.740 &  137.3775 & 3.6822 & 0.740 &  137.2023 & 3.5596 & 0.740 & 30.90 & 3.09 \\
102 & 81 & eFEDS J090930.6+034055--eFEDS J090718.6+035258 & 137.3775 & 3.6822 & 0.740 &  136.8278 & 3.8829 & 0.729 &  137.1026 & 3.7825 & 0.734 & 93.22 & 3.29 \\
103 & 83 & eFEDS J090700.7+011032--eFEDS J090418.6+020642 & 136.7531 & 1.1757 & 0.799 &  136.0778 & 2.1117 & 0.808 &  136.4155 & 1.6437 & 0.804 & 122.73 & 2.92 \\
104 & 83 & eFEDS J090418.6+020642--eFEDS J090821.9+025141 & 136.0778 & 2.1117 & 0.808 &  137.0912 & 2.8615 & 0.812 &  136.5845 & 2.4866 & 0.810 & 112.98 & 3.13 \\
105 & 83 & eFEDS J090452.4+033326--eFEDS J090033.7+033932 & 136.2187 & 3.5574 & 0.808 &  135.1408 & 3.6591 & 0.809 &  135.6797 & 3.6082 & 0.809 & 92.87 & 3.28 \\
106 & 84 & eFEDS J084035.8+044036--eFEDS J084016.6+033951 & 130.1493 & 4.6768 & 0.793 &  130.0695 & 3.6644 & 0.805 &  130.1094 & 4.1706 & 0.799 & 124.43 & 2.80 \\
\hline\hline
\end{longtable}
Column 1: filament ID. Column 2: ID of the super clusters. Columns 3, 4, and 5: RA, Dec of a paired cluster (in degrees) and its redshift. Column 6, 7, and 8: RA, Dec of the other paired cluster (in degrees) and its redshift. Column 9, 10, and 11: RA, Dec of the inter-cluster filament (in degrees) and its redshift. Column 12 and 13: length (i.e. the distance between the cluster centres minus their virial radii) and width (i.e. the mean of the virial radii of the paired clusters) of the filament (in Mpc).
\end{landscape}

\bsp	
\label{lastpage}
\end{document}